\documentclass{ceab}   %% loading ceab.cls creates CEAB style
\usepackage{epsfig}     %\  Used to include figures
\usepackage{graphicx}   %/
\usepackage{natbib}     % for producing References section 
\usepackage[T1]{fontenc} % for producing Polish vowels 
\usepackage{amsmath}
\def\runningtitle{Rodolfo Barb\'a Meeting}   
\def\articlenumber{7}
\def\ppages{1--4}
\usepackage[]{xcolor} % Pour les couleurs !
\definecolor{DeepPink3}{rgb}{0.86, 0.00, 0.45}
\definecolor{DeepSkyBlue3}{rgb}{0.00, 0.58, 0.71}
\definecolor{Green1}{rgb}{0.00, 0.8, 0.1}
\usepackage{pgf,tikz}

\begin{document}

\title{Apsidal motion in massive binaries -- or how to sound stellar interiors without asteroseismology}

\author{S. Rosu$^{1,2}$
\vspace{2mm}\\
\it $^1$ Observatoire de Genève, Chemin Pegasi 51, 1290 Versoix, Switzerland\\ 
\it E-mail: sophie.rosu@unige.ch\\
\it $^2$ Department of Physics, KTH Royal Institute of Technology, \\
\it The Oskar Klein Centre, AlbaNova, SE-106 91 Stockholm, Sweden
}

\maketitle

\begin{abstract}
One of the most efficient and reliable observational technique allowing to probe the internal structure of a star is the determination of the apsidal motion in close eccentric binary systems.\\
This secular precession of the binary orbit’s major axis depends on the tidal interactions occurring between the two stars. The rate of this motion is directly related to the internal structure of the stars, in particular their inner density profile.\\
Combining radial velocity and light curve measurements made over a long timescale, the rate of apsidal motion can be constrained, together with the fundamental parameters of the stars. The confrontation of the observationally determined parameters to theoretical models of stellar structure and evolution then allows us to constrain the internal structure of the stars.\\
This powerful technique has been known for years but has been seldom applied to massive stars. I will highlight its interest and reveal recent results concerning several massive binaries.\\
While standard 1D stellar evolution models predict stars having a smaller internal stellar structure constant, that is to say, stars having a smaller density contrast, than expected from observations, I demonstrate that the addition of mixing inside the models helps to solve, at least partially, this discrepancy. Whether this additional mixing might be fully explained by rotationally-induced mixing is under investigation. Studies with the non-perturbative code MoBiDICT showed that the perturbative model assumption is not justified in highly distorted stars, in which cases the apsidal motion is underestimated, exacerbating even more the need for enhanced mixing inside the models.\\
But what happens if the binary is a double-line spectroscopic but non-eclipsing binary? In that case, we indeed have no estimate of the masses and radii of the stars. Surprisingly, the apsidal motion equations combined with the binary’ spectroscopic observations allow us to derive the masses of the stars, in a model-dependent way. Rodolfo Barb\'a contributed to the development of this original method that I bring out.  
\end{abstract}


\section{The importance of apsidal motion in close eccentric (massive) binaries}
A very naive test one can do provided an assembly of astronomers is given consists in asking the assembly how to sound the stellar interiors. Very quickly, one will hear the answer "asteroseismology" (or an abstruse sound if the audience is likely to joke). In any case, chances are small one will get "apsidal motion" as an answer. Yet, to probe stellar interiors, the apsidal motion turns out to be a powerful technique, especially for the O-type stars for which asteroseismology has trouble to provide us with accurate and reliable results. In this paper, I demonstrate why and how the apsidal motion in close eccentric binaries is key to sound the interior of stars, with a specific honour to Rodolfo Barbá's important contribution in this field. 

\begin{figure}[b!]
\centering
\begin{tikzpicture}[scale=0.8]
\fill[blue] plot [smooth,domain=0:4] (\x,{sqrt(4-\x^2/4)});
\fill[gray!40](0,0) circle (4 and 2);
\draw (0,0) circle (4 and 2);
\fill[yellow!60,opacity=0.5,rotate=45](0,0) circle (4 and 2);
\draw[thick,rotate=45] (0,0) circle (4 and 2);
\fill[gray!40] (0.805,-1.955) -- (-0.804,1.955) -- plot (-0.804,1.955) [rotate=45] arc(78.3:140.3:4 and 2);
\draw[very thick, gray!40] plot (-0.815,1.94) [rotate=45] arc(78.40:139.9:4 and 2);
\fill[gray!40] plot (-3.06,-1.274)  arc(219.6:281.6:4 and 2);
\draw[gray!40,very thick] (-0.825,1.955) [rotate=45] arc(77.9:140:4 and 2);
\draw[Green1,thick,->](0.2,0)arc(0:300:0.2);
\draw (-4,0) arc (180:300:4 and 2);
\draw[Green1] (0.804,-1.955)--(-0.804,1.955);
\draw (-0.804,1.955) arc (101.5:105:4 and 2);
\draw[->, -latex](0,0) -- (5,0);
\draw[thick,red,rotate=45] (0,0) -- (4,0);
\draw[thick,red,rotate=45] (-2.55,0) -- (-4,0);
\draw[thick,red, ->](0.18, -0.4)arc(-60:45:0.4);
\draw[red](0.25,-0.3) node[right]{$\omega$};
\draw[thick,DeepSkyBlue3](0,0)--(0.5,2.8);
\draw[thick,DeepSkyBlue3, ->](0.4, 0.4)arc(45:90:0.4);
\draw[DeepSkyBlue3](0.35,0.41) node[above]{$\phi$};
\draw[DeepSkyBlue3](0.5,2.8)node{\huge{$\star$}};
\draw[DeepSkyBlue3](0.5,2.8)node[above left]{$P$};
\draw[DeepPink3,thick, ->](1.3,-1.88)arc(-70:60:0.2);
\draw[DeepPink3](1.35,-1.6)node[right]{$i$};
\draw[Green1](0.804,-1.955)node{$\bullet$};
\draw[Green1](0.9,-1.955)node[below]{$\Omega_0$};
\draw[Green1](-0.2,0)node[left]{$\Omega$};
\draw[red](0.3,0.35)node[below right,rotate=45]{\footnotesize{Line of apsides}};
\draw[red,rotate=45](4,0)node{$\bullet$};
\draw[red,rotate=45](-4,0)node{$\bullet$};
\draw[red,rotate=45](4,0)node[right]{\footnotesize{Periastron}};
\draw[red,rotate=45](-4,0)node[left]{\footnotesize{Apastron}};
\draw(-2,0)node{\footnotesize{Reference plane}};
\draw(-2,-0.3)node{\footnotesize{Plane of the sky}};
\draw(5,0)node[below]{\footnotesize{Reference}};
\draw(5,-0.3)node[below]{\footnotesize{direction}};
\draw[rotate=45](-2.8,-0.5)node[rotate=45]{\footnotesize{Orbital}};
\draw[rotate=45](-2.8,-0.8)node[rotate=45]{\footnotesize{plane}};
\draw[Green1](-0.7,1.05)node[rotate=-70]{\footnotesize{Line of nodes}};
\end{tikzpicture}
\caption{Definition of the orbital elements of a binary system. The argument of periastron, $\omega$, is the angle between the line of nodes and the line of apsides, and the true anomaly, $\phi$, is the angle between the line of apsides and the position of the primary star; Both are measured in the orbital plane.\label{fig:orbital}}
\end{figure}
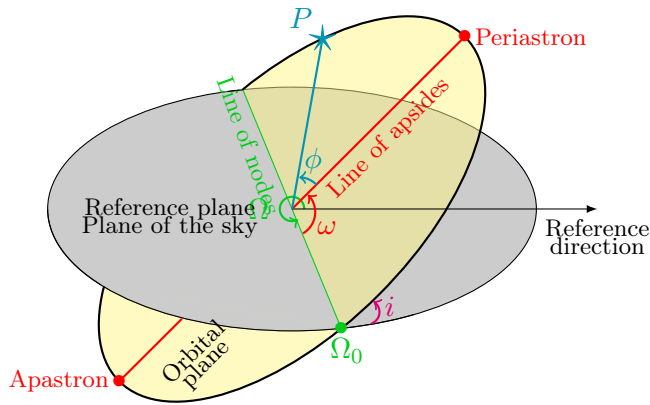

In celestial mechanics, the two-body Keplerian problem states that the two stars of a binary, assumed to be point-like particles, orbit each other on an elliptic orbit which orientation is fixed in space and time. The position of the stars on their orbit, measured through the true anomaly (see Fig.\,\ref{fig:orbital}) is the only time-varying parameter. While the Keplerian problem is arguably valid for wide eccentric binaries, it is inappropriate for close eccentric binaries. Indeed, the tidal interactions occurring between the two stars are responsible for the non-spherical gravitational fields of the stars consequent to the stellar deformations and for the exchanges of angular momentum in the binary. The orbital motion is directly affected: The line of apsides (the line joining the periastron and apastron, see Fig.\,\ref{fig:orbital}) precesses in time, a motion known as the apsidal motion. This motion, usually on the order of a few degrees per year, is purely Newtonian and should not be mistaken with the general relativistic contribution to the apsidal motion -- of minor importance for most binaries. 

\begin{figure}[t!]
\centering
\includegraphics[clip=true,trim=10 10 10 10,width=0.45\linewidth]{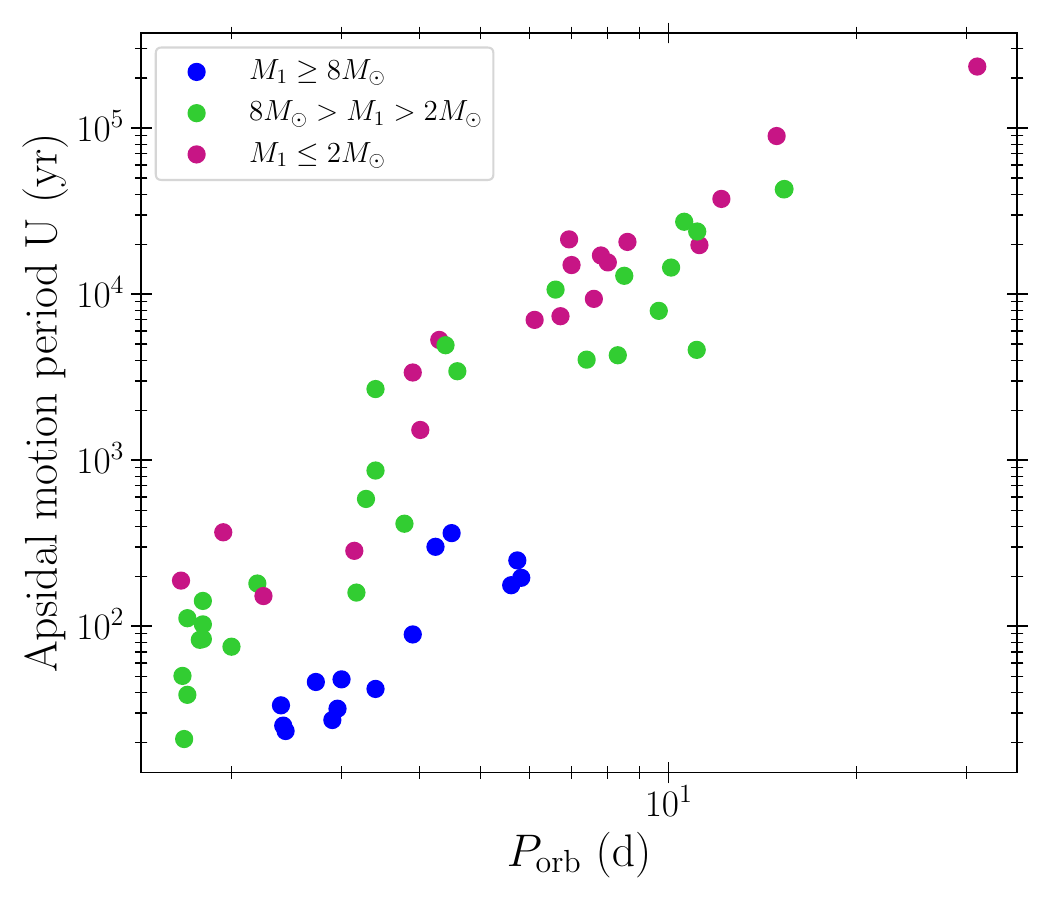}\\
\includegraphics[clip=true,trim=10 0 0 10,width=0.45\linewidth]{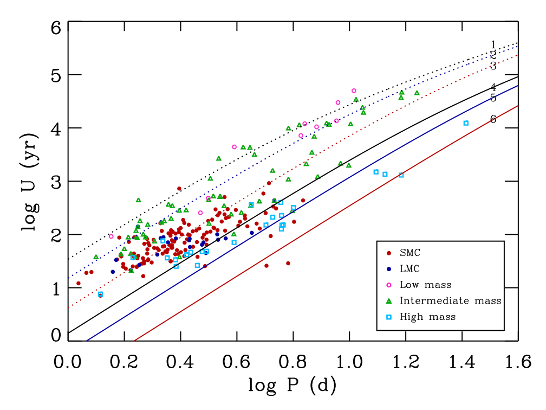}
\includegraphics[clip=true,trim=10 0 10 10,width=0.53\linewidth]{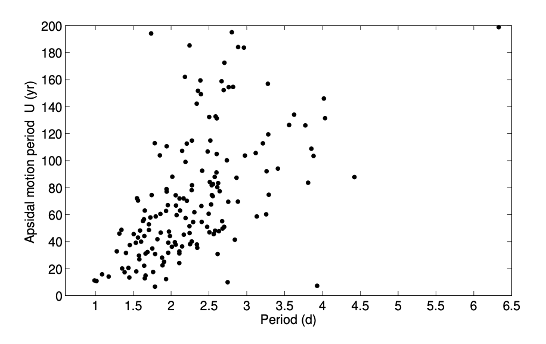}
\bigskip
\begin{minipage}{12cm}
\caption{Apsidal motion period as a function of the orbital period. \textit{Upper panel:} individual observed systems coming from the literature \citep{baroch21, baroch22, claret21, marcussen22, rauw16, rosu20b, rosu22a, rosu22b, rosu23, torres10, wolf06, wolf08, wolf10}. Only the primaries are plotted, colour-coded by their mass. Figure taken from \citet{rosu24b}. \textit{Lower left panel:} study of \citet{hong16}. Credit: Figure 6 of \citet{hong16}. \textit{Lower right panel:} study of \citet{zasche19, zasche20}. Credit: \citet{zasche20}, reproduced with permission \copyright ESO.}
\label{fig:catalogue}
\end{minipage}
\end{figure}

Though relatively small, the apsidal motion has been observed and measured in hundredth of systems (see Fig.\,\ref{fig:catalogue}). Traditionally, the apsidal motion rate determination from photometric observations of eclipsing binaries is favoured over the determination from radial velocities (RVs) of spectroscopic binaries. The high-quality space-borne Kepler and TESS data indeed allow the accurate determination of the apsidal motion rate using a much shorter time span of the observations than the decades of observations necessary for the RVs adjustments. Regardless, when sufficient spectroscopic observations are available, this latter technique turns out to be very powerful and more accurate. It goes without saying that short-period eccentric, double-line spectroscopic, eclipsing massive binaries for which spectroscopic and photometric observations are available are intrinsically invaluable systems to derive the apsidal motion together with all stellar and orbital parameters in a model-independent way. 

\subsection{Eclipsing binary}
\begin{figure}[t!]
\centering
\includegraphics[clip=true,trim=0 0 0 0,width=0.5\linewidth]{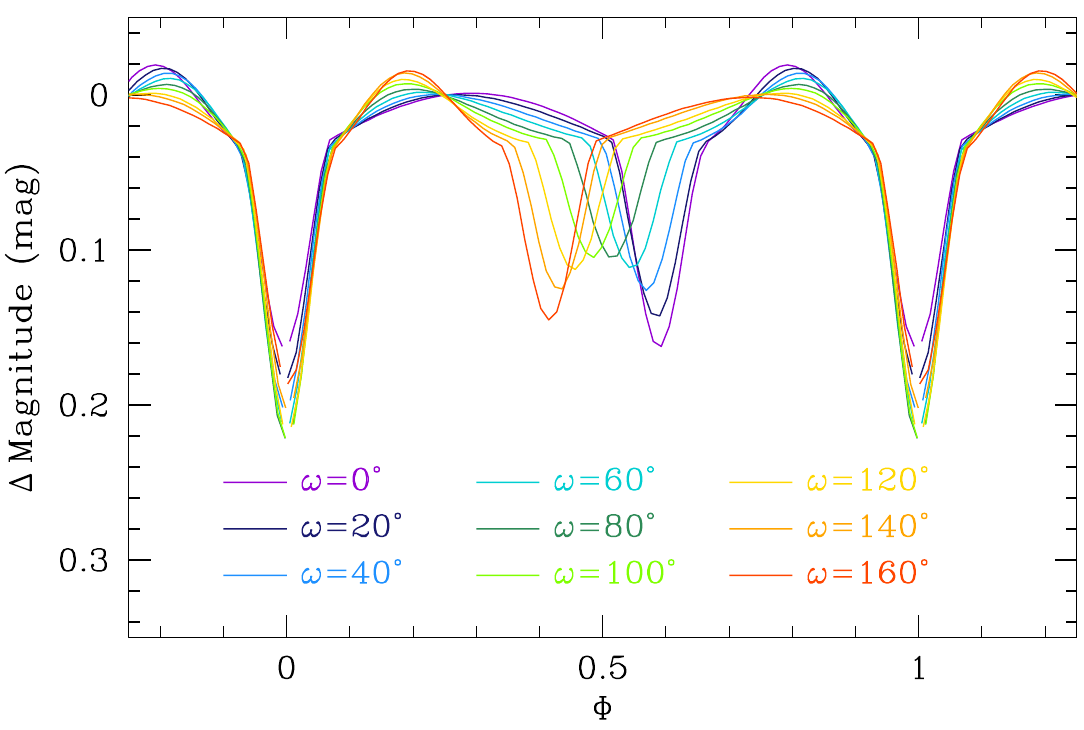}
\caption{Nine theoretical lightcurves colour-coded by the assumed value of $\omega$. All other parameters are identical ($e=0.134$, $i=68.6^\circ$, and the stars have the same $m, R$, and $T_\mathrm{eff}$). Figure from \citet{rosu21}.}
\label{fig:lightcurves}
\end{figure}

\begin{figure}[t!]
\centering
\includegraphics[clip=true,trim=10 8 10 10,width=0.67\linewidth]{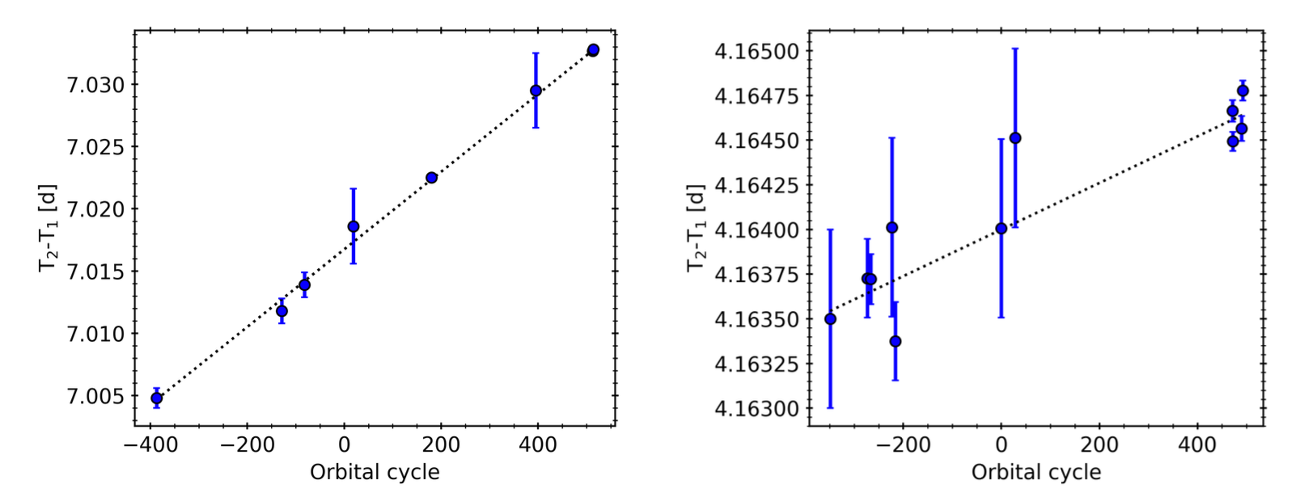}
\includegraphics[clip=true,trim=30 35 10 130,width=0.32\linewidth]{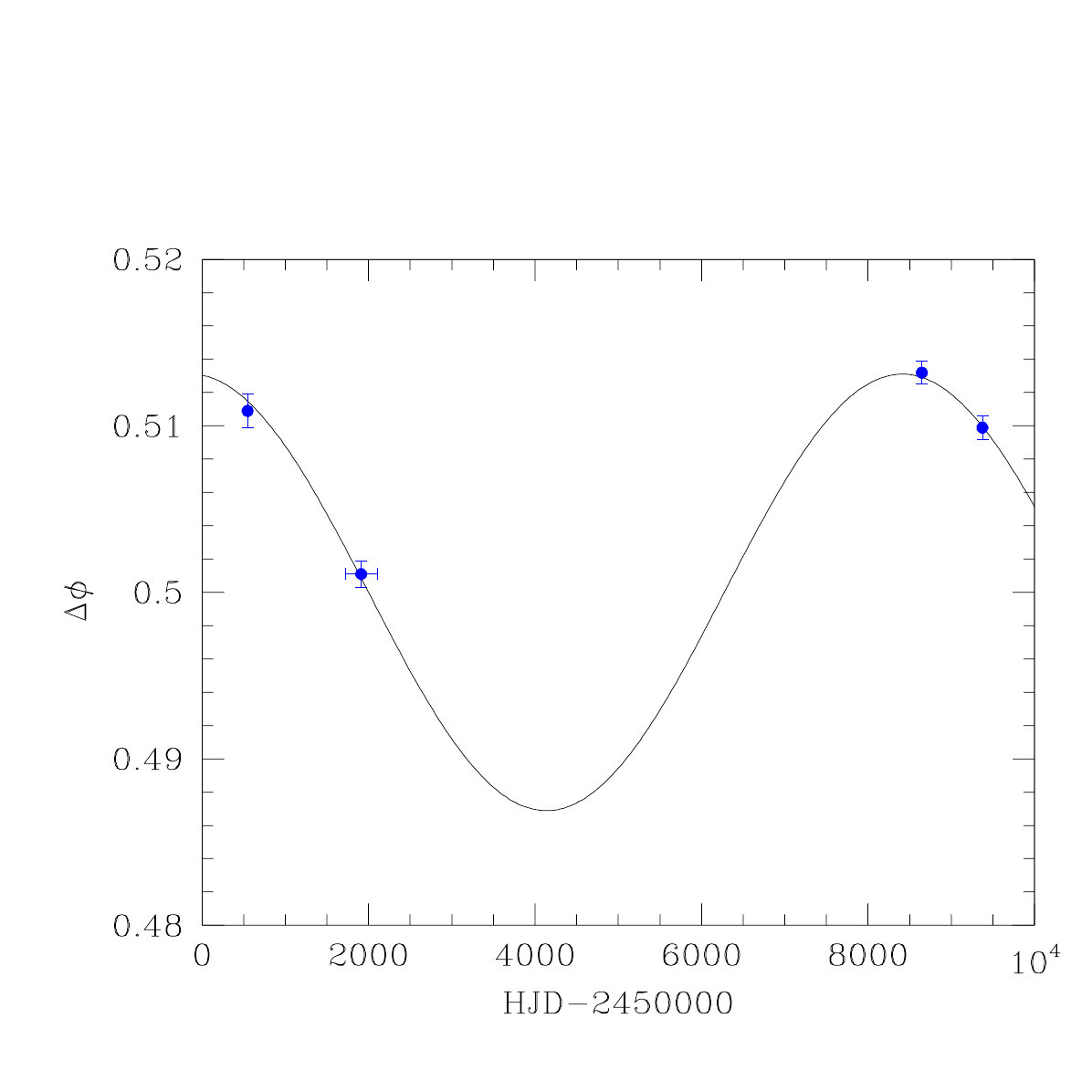}
\caption{\textit{Left panel:} Time difference between secondary and primary minima of the eclipses as a function of the orbital cycle for V541\,Cyg. Figure from \citet{baroch21}. \textit{Middle panel:} Same for V459\,Cas. Figure from \citet{baroch21}. \textit{Right panel:} Phase difference between the secondary and primary minima as a function of time for CPD-41$\circ$\,7742. Figure from \citet{rosu22b}.}
\label{fig:times_minima}
\end{figure}

The effect of a change in the longitude of periastron $\omega$ on the lightcurves of an eclipsing binary is schematically illustrated in Fig.\,\ref{fig:lightcurves}. For this example, we adopted a twin binary (meaning that both stars share the same properties in terms of mass $m$, radius $R$, and effective temperature $T_\text{eff}$) and different values of $\omega$ ranging from $0^\circ$ to $160^\circ$. For these values of $\omega$, the primary (resp. secondary) eclipse happens closer to periastron (resp. apastron) passage; It explains why the primary eclipse is systematically deeper than the secondary eclipse. Because the binary is a twin, the difference in depth between the two eclipses is not related to the effective temperatures of the stars, but is entirely attributable to the combined effect of the inclination of the system, the eccentricity of the orbit, and the corresponding $\omega$. The change in depth and phase of the eclipses follow the change in $\omega$: They both depend upon the orbital separation at conjunction phase which itself depends upon the orientation of the ellipse with respect to our line of sight, that is to say, $\omega$. The two methods to derive the apsidal motion rate based on a set of photometric data are 1) the fit of the lightcurves of a binary taken at different epochs with $\omega$ as a free parameter and 2) the fit of the times of minima of the eclipses using the equations of \citet{gimenez95} to fit the phase differences ($\Delta\phi$) between the primary and secondary minima ($T_1$ and $T_2$) at different epochs. The latter method is illustrated for the binaries V541\,Cyg and V459\,Cas \citep{baroch21}, and CPD-41$^\circ$\,7742 \citep{rosu22b} in Fig.\,\ref{fig:times_minima}. The apsidal motion rates amount to respectively $0.0084\pm0.0001^\circ$\,yr$^{-1}$, $0.028\pm0.004^\circ$\,yr$^{-1}$, and $15.38\pm0.51^\circ$\,yr$^{-1}$.

\subsection{Spectroscopic binary}
\begin{figure}[h]
\centering
\includegraphics[clip=true,trim=0 200 0 0,width=0.7\linewidth]{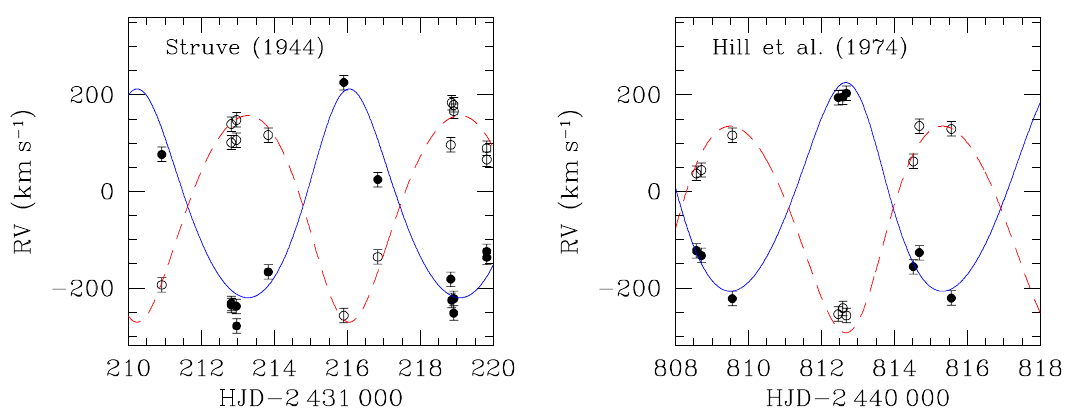}
\includegraphics[clip=true,trim=0 0 0 200,width=0.7\linewidth]{S_ROSU_FIG5-eps-converted-to.pdf}
\caption{Measured RVs of the primary (filled dots) and secondary (open dots) stars of HD\,152248, and best-fit RV curves (blue and red). Data from \citet{struve44, hill74, penny99, rosu20b}. Figure from \citet{rosu21}.}
\label{fig:RVfit}
\end{figure}

In a double-line spectroscopic binary, the apsidal motion rate is derived from the fit of a set of primary (P) and secondary (S) RVs according to the following equations
\begin{equation}
\label{eqn:RVp}
RV_\text{P}(t) = K_\text{P}\left(\cos(\phi(t)+\omega(t))+e\cos(\omega(t))\right) + \gamma_\text{P},\\
\end{equation}
\begin{equation}
\label{eqn:RVs}
RV_\text{S}(t) = -K_\text{S}\left(\cos(\phi(t)+\omega(t))+e\cos(\omega(t))\right) + \gamma_\text{S},\\
\end{equation}
accounting for the apsidal motion rate $\dot\omega$ through the linear change of $\omega$ with time: 
\begin{equation}
\omega(t) = \omega_0 + \dot\omega(t-T_0),
\end{equation}
where $\omega_0$ is the value of $\omega$ at the reference time $T_0$. In Eqs\,\eqref{eqn:RVp} and \eqref{eqn:RVs}, $K_*$ and $\gamma_*$ are the semi-amplitude of the RV curve and the apparent systemic velocity of the corresponding star. An example of RVs adjustments is presented in Fig.\,\ref{fig:RVfit} \citep[see][for further examples]{rauw16, rosu22a, rosu22b, rosu23, barclay24}. 

\subsection{Eclipsing and spectroscopic binary}
\begin{figure}[h]
\centering
\includegraphics[clip=true,trim=20 200 20 20,width=0.6\linewidth]{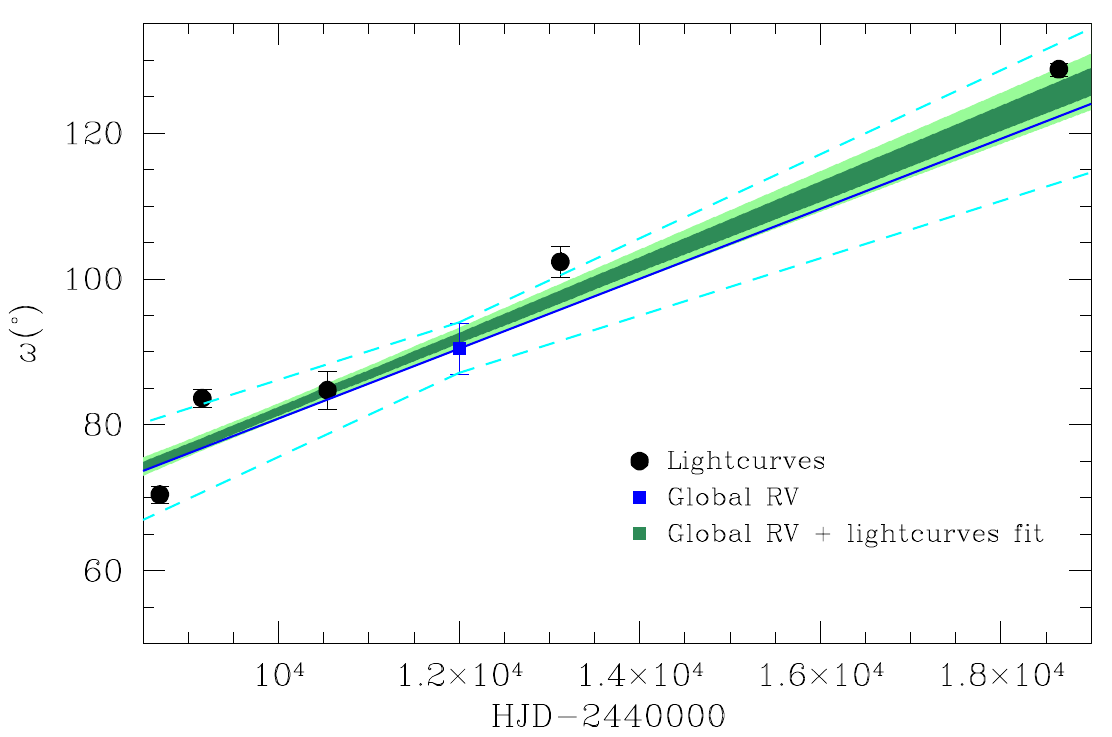}
\caption{Values of $\omega$ as a function of time inferred from the RVs and lightcurves of HD\,152248. The $\omega_0$ value obtained from the fit of all RVs is the blue filled square. The best value of $\dot\omega$ and its $1\sigma$ uncertainties from the RVs are the solid blue line and the dashed cyan lines. Each individual fit of a lightcurve is represented by a black dot. The $1\sigma$ and $2\sigma$ uncertainties on $\omega$ from the simultaneous fit of all data with \texttt{PHOEBE} are in dark and light green. Credit: \citet{rosu20b}, reproduced with permission \copyright ESO.}
\label{fig:HD152248_ajustement}
\end{figure}

\begin{figure}[h!]
\centering
\includegraphics[clip=true,trim=0 0 0 0,width=0.44\linewidth]{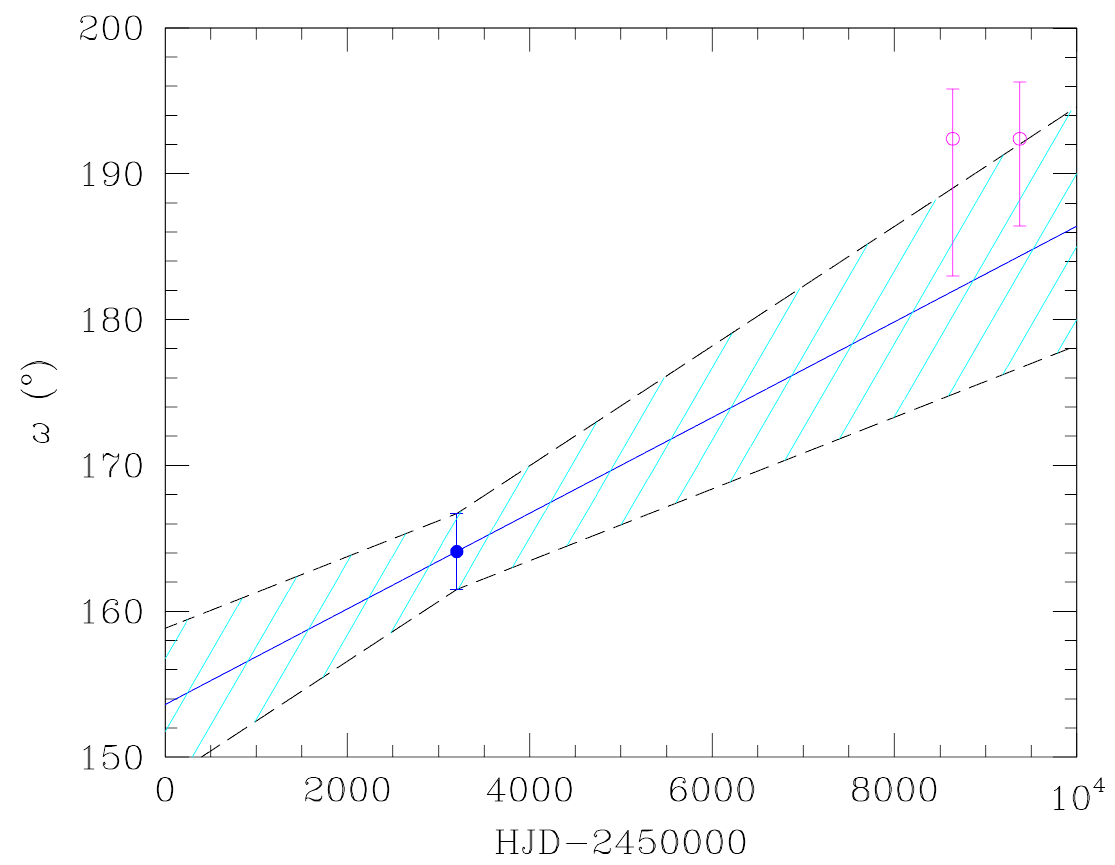}
\includegraphics[clip=true,trim=0 0 0 0,width=0.44\linewidth]{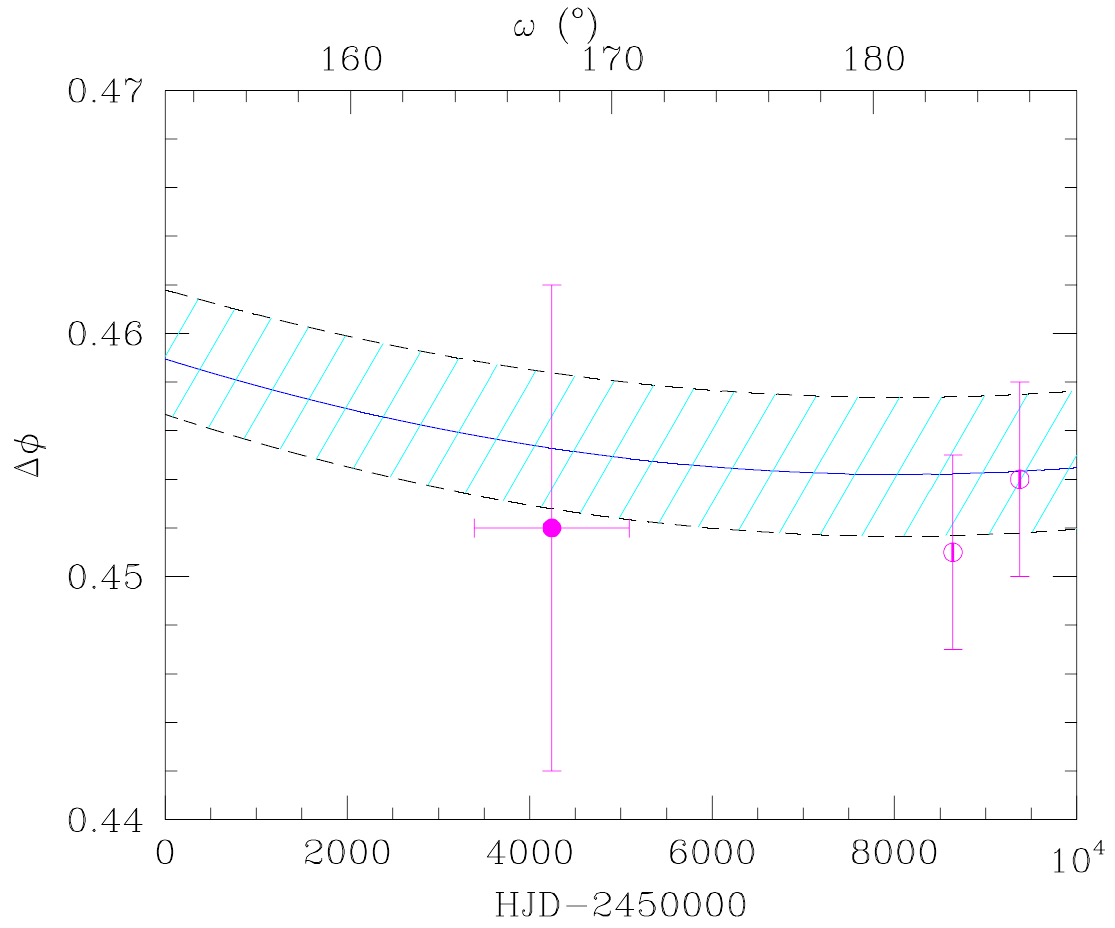}
\caption{\textit{Left panel:} Values of $\omega$ as a function of time inferred from the RVs and lightcurves of HD\,152219. The fits of the photometry are in pink. The $\omega_0$ value obtained from the global fit of all RV data is the blue dot, while the solid blue line corresponds to the best-fit value of $\dot\omega$ inferred from the RVs, and the hatched cyan zone corresponds to the range of values according to the $1\sigma$ uncertainties on $\omega_0$ and $\dot\omega$. \textit{Right panel:} Values of the phase difference $\Delta\phi$ between the primary and secondary eclipses as a function of time and $\omega$ inferred from the RVs and lightcurves. The fits of the photometry are in pink. The solid blue line corresponds to the best-fit value of $\Delta\phi$  inferred from the RVs, and the hatched cyan zone corresponds to the range of values according to the $1\sigma$ uncertainties on $e$. Figures from \citet{rosu22a}.}
\label{fig:HD152219_ajustement}
\end{figure}

We illustrate in Figs\,\ref{fig:HD152248_ajustement} and \ref{fig:HD152219_ajustement} the cases of two eclipsing and double-line spectroscopic massive binaries, HD\,152248 and HD\,152219, for which an apsidal motion rate was derived based on the combined analysis of the spectroscopic and photometric observations of the systems. The apsidal motion rates amount to $1.84\pm0.08^\circ$\,yr$^{-1}$ and $1.20\pm0.30^\circ$\,yr$^{-1}$, respectively.

\section{Apsidal motion as a means to sound stellar interiors} 
\begin{figure}[t!]
\centering
\includegraphics[clip=true,trim=40 20 150 100,width=0.49\linewidth]{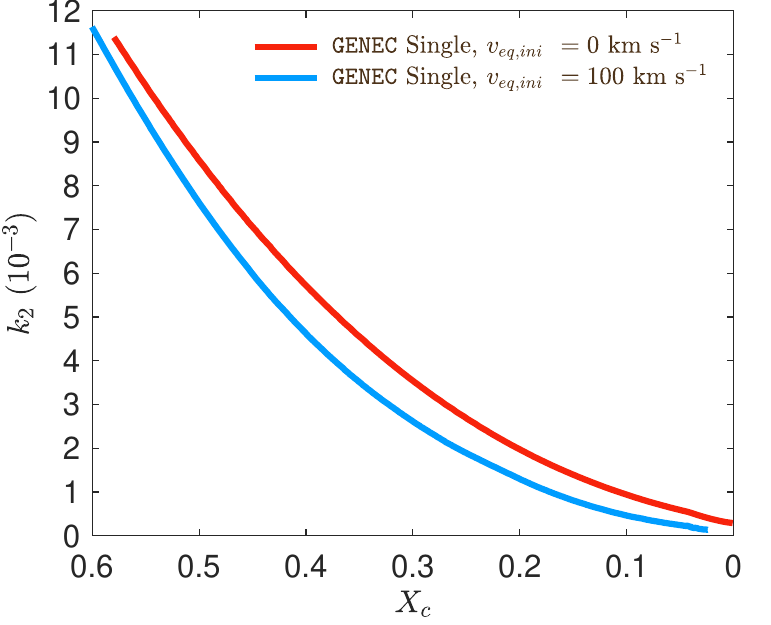}
\includegraphics[clip=true,trim=40 20 150 100,width=0.49\linewidth]{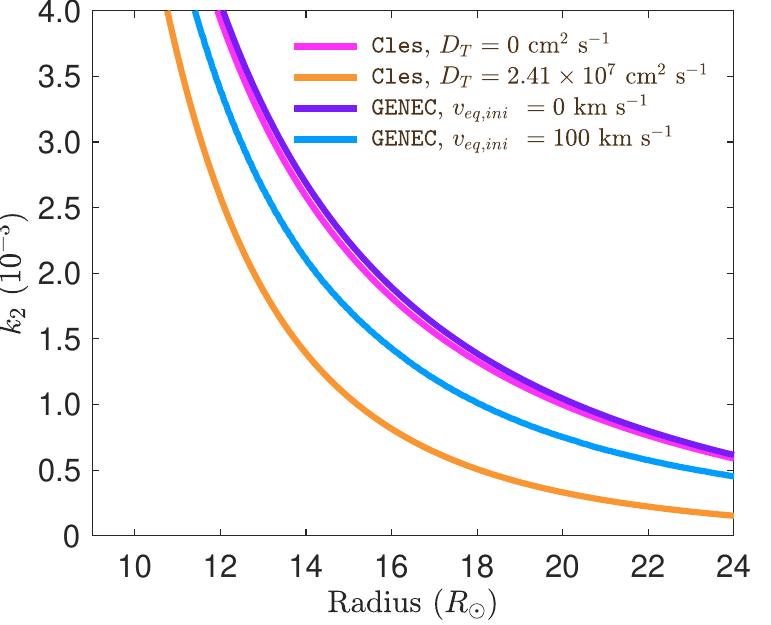}
\caption{Evolution of $k_2$ as a function of the hydrogen mass fraction $X_c$ inside the star (\textit{left panel}) and the radius of the star (\textit{right panel}) for \texttt{GENEC} models with or without initial rotational velocity and \texttt{Cl\'es} models with or without turbulent diffusion. All models have an initial mass of $32.8\,\text{M}_\odot$. Credit: \citet{rosu20a}, reproduced with permission \copyright ESO.}
\label{fig:k2XcR}
\end{figure}

In the perturbative case, the apsidal motion rate consists in the Newtonian contribution (N) plus the general relativistic correction (GR): 
\begin{equation}
\label{eqn:omegadot} 
\dot\omega = \dot\omega_\text{N} + \dot\omega_\text{GR}. 
\end{equation}
Their expressions are given by 
\begin{equation}
\label{eqn:omegadotN}
\begin{aligned}
\dot\omega_\mathrm{N} = &\frac{2\pi}{P_\text{orb}} \Bigg[15f(e)\left\{k_{2,1}q \left(\frac{R_1}{a}\right)^5 + \frac{k_{2,2}}{q} \left(\frac{R_2}{a}\right)^5\right\} \\
& + g(e) \Bigg\{ k_{2,1} (1+q) \left(\frac{R_1}{a}\right)^5 \left(\frac{P_\mathrm{orb}}{P_\text{rot,1}}\right)^2+ k_{2,2}\, \frac{1+q}{q} \left(\frac{R_2}{a}\right)^5 \left(\frac{P_\mathrm{orb}}{P_\text{rot,2}}\right)^2 \Bigg\}  \Bigg],
\end{aligned}
\end{equation}
where only the contributions arising from the second-order harmonic distorsions of the gravitational potential are considered \citep{sterne39}, and by
\begin{equation} 
\label{eqn:omegadotGR}
\dot\omega_\mathrm{GR}  = \left(\frac{2\pi}{P_\mathrm{orb}}\right)^{5/3}\frac{3(G(m_1+m_2))^{2/3}}{c^2 (1-e^2)}
\end{equation}
\citep{shakura85}. In these expressions, $q=m_2/m_1$ is the mass ratio, $P_\text{orb}$ is the orbital period of the system, $a$ is the semi-major axis of the orbit, $R_*$ and $P_{\mathrm{rot,}*}$ are the radius and the rotational period of the considered star, $f(e)$ and $g(e)$ are functions of the eccentricity of the orbit which expressions are given in e.g., \citet{rosu21}, $G$ is the gravitational constant, and $c$ is the speed of light.\\
In Eq.\,\eqref{eqn:omegadotN}, $k_{2,*}$ is the internal structure constant of the considered star:
\begin{equation}
\label{eqn:k2eta}
k_2 = \frac{3-\eta_2(R_{*})}{4+2\,\eta_2(R_{*})},
\end{equation}
where $\eta_2(R_*)$ is the solution evaluated at the stellar surface of the Clairaut-Radau differential equation \citep{hejlesen87}:
\begin{equation}
r \frac{d\eta_2(r)}{dr} + \eta_2^2(r) - \eta_2(r) + 6 \frac{\rho(r)}{\bar\rho(r)} \left(\eta_2(r)+1\right) - 6 = 0,
\label{eqn:Radau}
\end{equation}
with the boundary condition $\eta_2(0) = 0$, where $\rho$ is the density in a shell at a distance $r$ from the centre and $\bar\rho$ is the mean density inside the sphere of radius $r$. 
Hence, $k_2$ depends upon the density profile inside the star. More precisely, $k_2$ measures the density stratification between the core and the external layers of the star. $k_2$ takes its maximum value of 0.75 for an homogeneous sphere of constant density but takes values as low as $10^{-4}$ for massive stars which have a dense core and a diluted envelope \citep{rosu20a}. As its core contracts and envelope expends during its evolution, the star sees its $k_2$ decreases in time, rendering $k_2$ a good indicator of stellar evolution. This is better illustrated in Fig.\,\ref{fig:k2XcR} where the evolution of $k_2$ is presented as a function of the hydrogen mass fraction inside the star -- the main indicator of stellar evolution -- (left) as well as as a function of the stellar radius (right) for several \texttt{GENEC}\footnote{\texttt{GENEC} is developed and maintained at the Geneva Observatory, Switzerland, see e.g., \citet{eggenberger08}.} and \texttt{Cl\'es}\footnote{\texttt{Cl\'es} is developed and maintained at the University of Li\`ege, Belgium, see e.g., \citet{scuflaire08}.} models. 

In the apsidal motion equations, all terms -- except for $k_{2,1}$ and $k_{2,2}$ -- can be derived from observations provided the binary is an eclipsing double line spectroscopic system. Only in the case of a twin system can the equations be solved for $k_2$ as $k_{2,1} = k_{2,2}$ in that case. In the general case, the system is underdetermined, but a weighted-averaged mean of the $k_2$ of the two stars can be obtained: 
\begin{equation}
\bar{k}_2 = \frac{c_1k_{2,1}+c_2k_{2,2}}{c_1+c_2},
\end{equation}
where the expressions of $c_1$ and $c_2$ are evident if we rewrite Eq.\,\eqref{eqn:omegadotN} as 
\begin{equation}
\dot\omega_\text{N} = c_1k_{2,1} + c_2k_{2,2}. 
\end{equation}
We refer to \citet{rosu22a} for a thorough discussion about how to still get information on the primary star in that general case.

\subsection{Enhanced mixing in the models as a necessity to reproduce the stellar density stratification}

\begin{figure}[b!]
\centering
\includegraphics[clip=true,trim=10 10 10 0,width=0.5\linewidth]{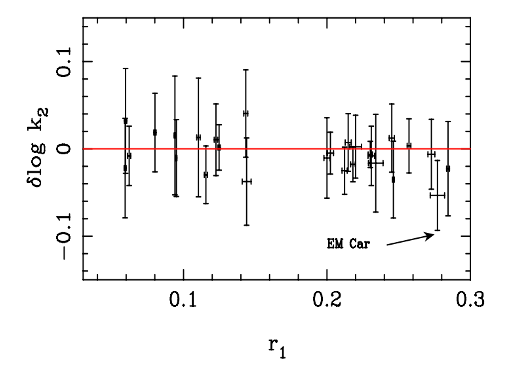}
\caption{Logarithmic difference between $k_{2,\text{obs}}$ and $k_{2,\text{th}}$ as a function of $r_1=R_1/a$ for 27 binaries. Credit: \citet{claret21}, reproduced with permission \copyright ESO.}
\label{fig:claret}
\end{figure}

\citet{claret21} confronted the theoretical (from stellar evolution models) and observational $k_2$ of 27 double-line eclipsing binaries with stellar mass ranging from 1.1 to 22.8\,M$_\odot$. The authors found a good agreement for systems which ratio $r_1 = R_1/a < 0.18$: $\log(k_{2,\text{obs}}) - \log(k_{2,\text{th}}) = -0.002 \pm 0.012$. But they found a systematic lower $k_{2,\text{obs}}$ for systems which $r_1 > 0.18$, that is to say, more massive and closer systems for which the tidal interactions are more pronounced: $\log(k_{2,\text{obs}}) - \log(k_{2,\text{th}}) = -0.010 \pm 0.008$, as shown in Fig.\,\ref{fig:claret}. 

\citet{rosu20a} and \citet{rosu22a} built dedicated stellar evolution models with \texttt{Cl\'es} for HD\,152248 and HD\,152219, respectively, with different prescriptions for the internal mixing. The  overshooting is implemented as a step-function with the parameter $\alpha_\text{ov}$, while additional mixing is introduced through the turbulent diffusion. The latter is a partial mixing process acting on the velocities of the chemical elements: $V_i = -D_\text{T} \frac{d\ln X_i}{dr}$ for element $i$, where $D_\text{T}$ is the turbulent diffusion coefficient, measured in cm$^2$\,s$^{-1}$. Turbulent diffusion acts as reducing the abundance gradient of the chemical elements as it brings additional hydrogen hence fuel for nuclear reactions from the star's external layers to the star's core. It increases the main-sequence lifetime of the star and as such, has a similar consequence as overshooting. 

\begin{figure}[t!]
\centering
\includegraphics[clip=true,trim=20 20 160 100,width=0.49\linewidth]{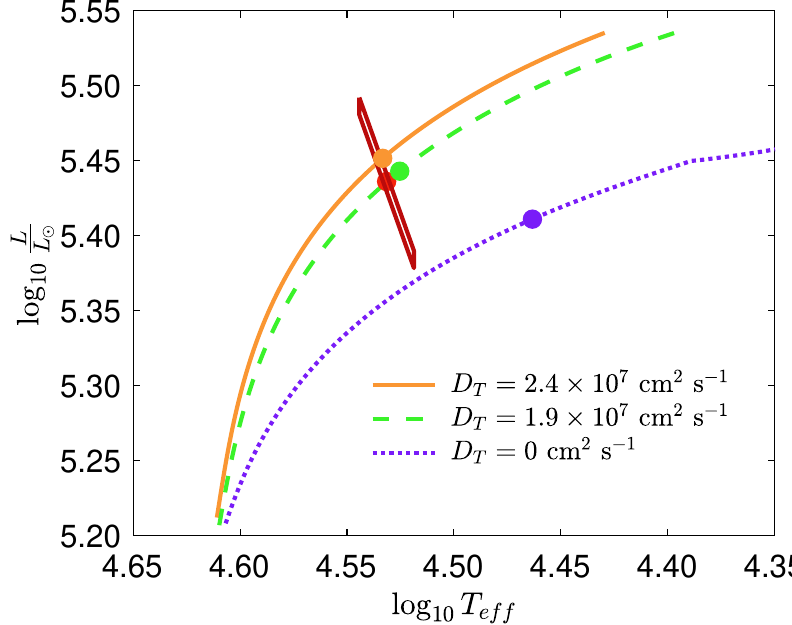}
\includegraphics[clip=true,trim=20 20 160 100,width=0.49\linewidth]{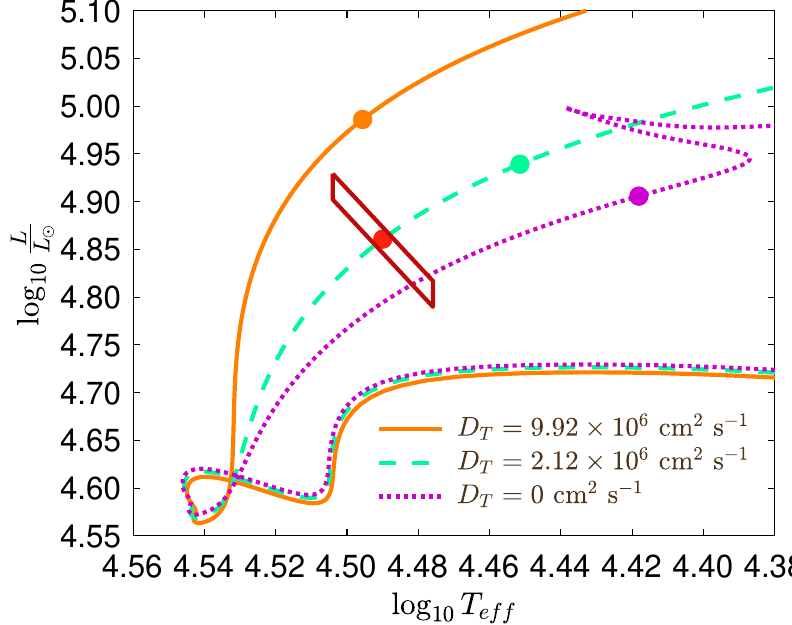}
\caption{HR diagrams for HD\,152248 (\textit{left panel}) and HD\,152219 (\textit{right panel}): evolutionary tracks of \texttt{Cl\'es} models. The observational values and their error bars are represented in red. The dots over-plotted on the tracks correspond to the models that fit the observational $k_2$. \textit{Left panel:} credit: \citet{rosu24c}, reproduced with permission \copyright ESO. \textit{Right panel:} figure adapted from \citet{rosu22a}.}
\label{fig:HR}
\end{figure}

The authors used the \texttt{min-Cl\'es} routine which implements the Levenberg-Marquardt minimisation technique to search for best-fit models of the stars in terms of observational properties (mass, radius, effective temperature, apsidal motion rate). Assuming as constraints the stellar mass, radius, and position in the Hertzsprung-Russell (HR) diagram only, and fixing $\alpha_\text{ov} = 0.20$ and $D_\text{T} = 0$\,cm$^2$\,s$^{-1}$, the authors could not find any model able to reproduce the physical properties of the stars (see the purple tracks in Fig.\,\ref{fig:HR}). In contrast, leaving $D_\text{T}$ as a free parameter of the adjustment, the authors found models that perfectly reproduce the mass, radius, and position in the HR diagram (see green tracks in Fig.\,\ref{fig:HR}). However, these best-fit models do not reproduce the $k_2$ of the stars. Indeed, the models that reproduce the $k_2$, shown by the dots overplotted on the stellar tracks, are located further away on the main sequence. The best-fit models in terms of the mass, radius, and position in the HR thus have a too-high $k_2$ value, hence a too low density contrast between the stellar core and external layers. The authors further required the models to reproduce not only the mass, radius, and position in the HR diagram, but also the apsidal motion rate through the $k_2$ and obtained best-fit models with enhanced turbulent diffusion (see orange tracks in Fig.\,\ref{fig:HR}). These models allow to solve, at least partially, the $k_2$ discrepancy. 

That stellar evolution models predict stars having a too low density stratification between their core and external layers compared to what observations suggest is a key result that could not be achieved without the study of the apsidal motion in massive binaries. The physical origin of this turbulent diffusion is yet to be determined: A preliminary study from \citet{rosu20a} suggests that rotationally-induced mixing could be the answer; Further investigations are ongoing with \texttt{GENEC}. 

\begin{figure}[h!]
\centering
\includegraphics[clip=true,trim=0 0 0 0,width=\linewidth]{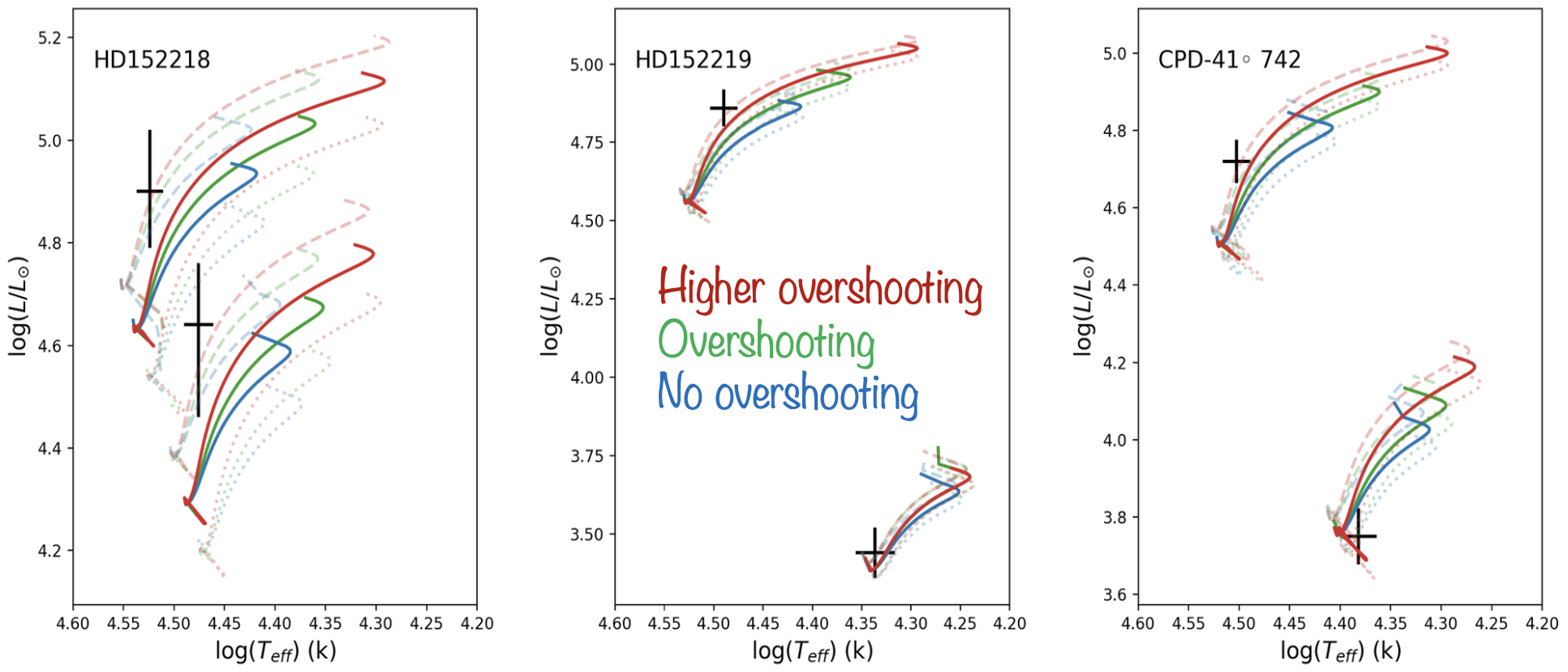}
\caption{HR diagrams for HD\,152218 (\textit{left panel}), HD\,152219 (\textit{middle panel}), and CPD-41$^\circ$\,7742 (\textit{right panel}): evolutionary tracks with different values of $\alpha_\text{ov}$. The observational values and their error bars are represented by the black crosses. Figures adapted from \citet{baraffe23}.}
\label{fig:baraffe}
\end{figure}

In another context, \citet{baraffe23} performed 2D hydrodynamical simulations to fit the stars of HD\,152219, HD\,152218, and CPD-41$^\circ$\,7742 and showed that a systematic large overshooting was necessary to reproduce the positions of the stars in the HR diagram (see Fig.\,\ref{fig:baraffe}). Based on similar models, \citet{baraffe23} showed that models without overshooting are unable to reproduce the main sequence width and that additional mixing is necessary, especially for higher masses (see their figure 7). These results are in agreement with those of \citet{castro14} and \citet{martinet21}.

\subsection{Going from 1D to 3D} 
The analyses of the apsidal motion performed so far suffer from one limitation: The perturbative model assumption adopted in the 1D stellar modelling considers the centrifugal and tidal forces as small perturbations of the spherical symmetry and only accounts for leading terms. \citet{fellay23} developed the code \texttt{MoBiDICT} (Modelling Binaries Deformations Induced by Centrifugal and Tidal forces), a non-perturbative method accounting for the entire precise 3D deformed structure of each component including the effects of stellar deformations on the mass redistribution. The instantaneous non-perturbative tidal acceleration perturbation and its consequence on the apsidal motion are calculated as explained in \citet{fellay23, fellay24}. 

\begin{figure}[h]
\centering
\includegraphics[clip=true,trim=10 10 10 10,width=0.49\linewidth]{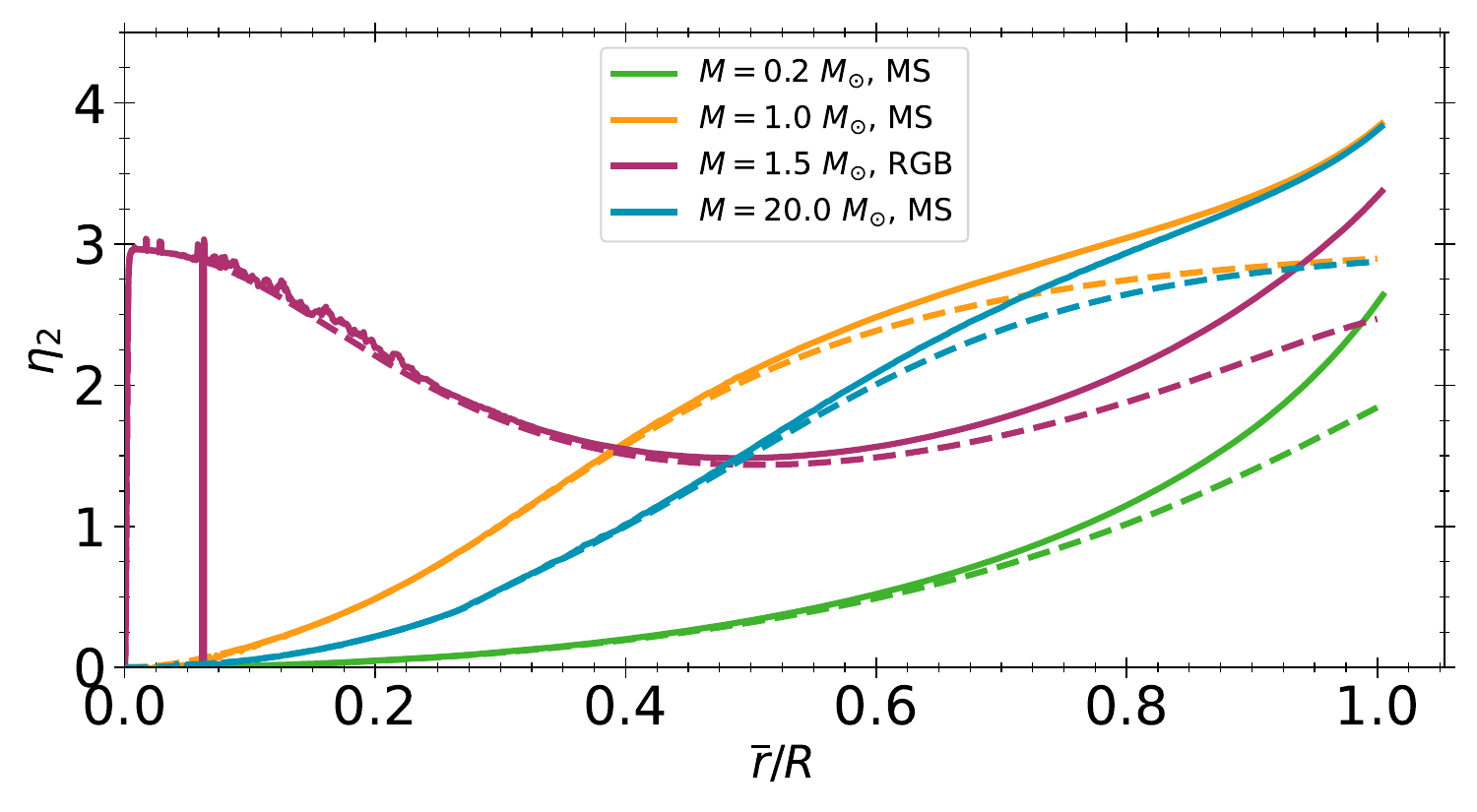}
\includegraphics[clip=true,trim=10 10 10 10,width=0.49\linewidth]{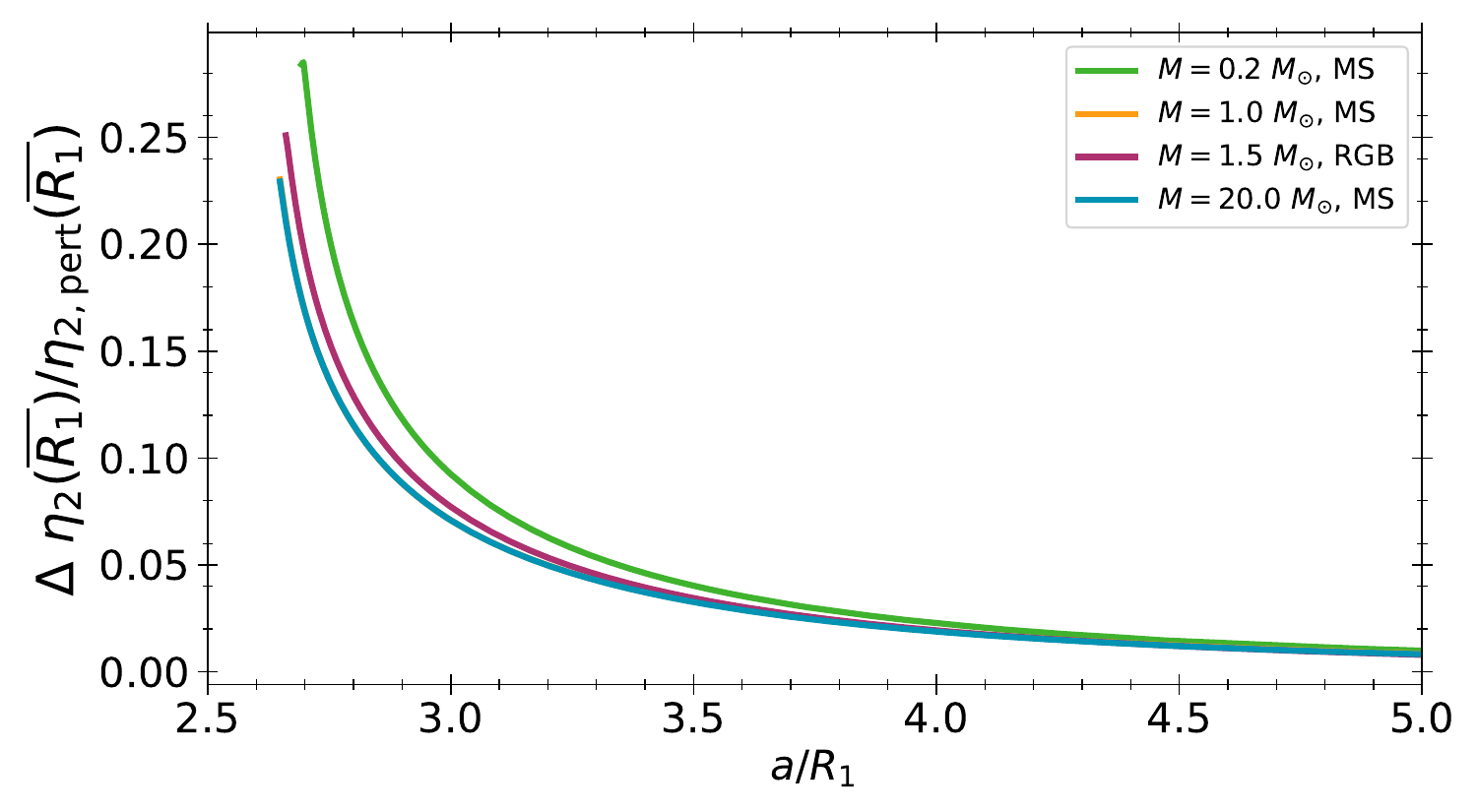}
\caption{\textit{Left panel:} $\eta_2(R_*)$ as a function of the radial distance to the core normalised by the radius for four different stars belonging to twin binaries made of two main-sequence stars of 0.2\,M$_\odot$ (green), 1.0\,M$_\odot$ (orange), or 20.0\,M$_\odot$ (blue) each, or two red giants of 1.5\,M$_\odot$ (purple) each, separated by a distance $a = 2.8\,R_1$. Solid and dashed lines correspond to non-perturbative and perturbative models, respectively. \textit{Right panel:} Relative difference between the $\eta_2(R_*)$ computed with \texttt{MoBiDICT} and with the perturbative approach as a function of the semi-major axis of the orbit scaled by the radius of the star for the same systems. Figures adapted from \citet{fellay23}.}
\label{fig:eta2}
\end{figure}

\begin{figure}[h!]
\centering
\includegraphics[clip=true,trim=10 10 10 10,width=0.49\linewidth]{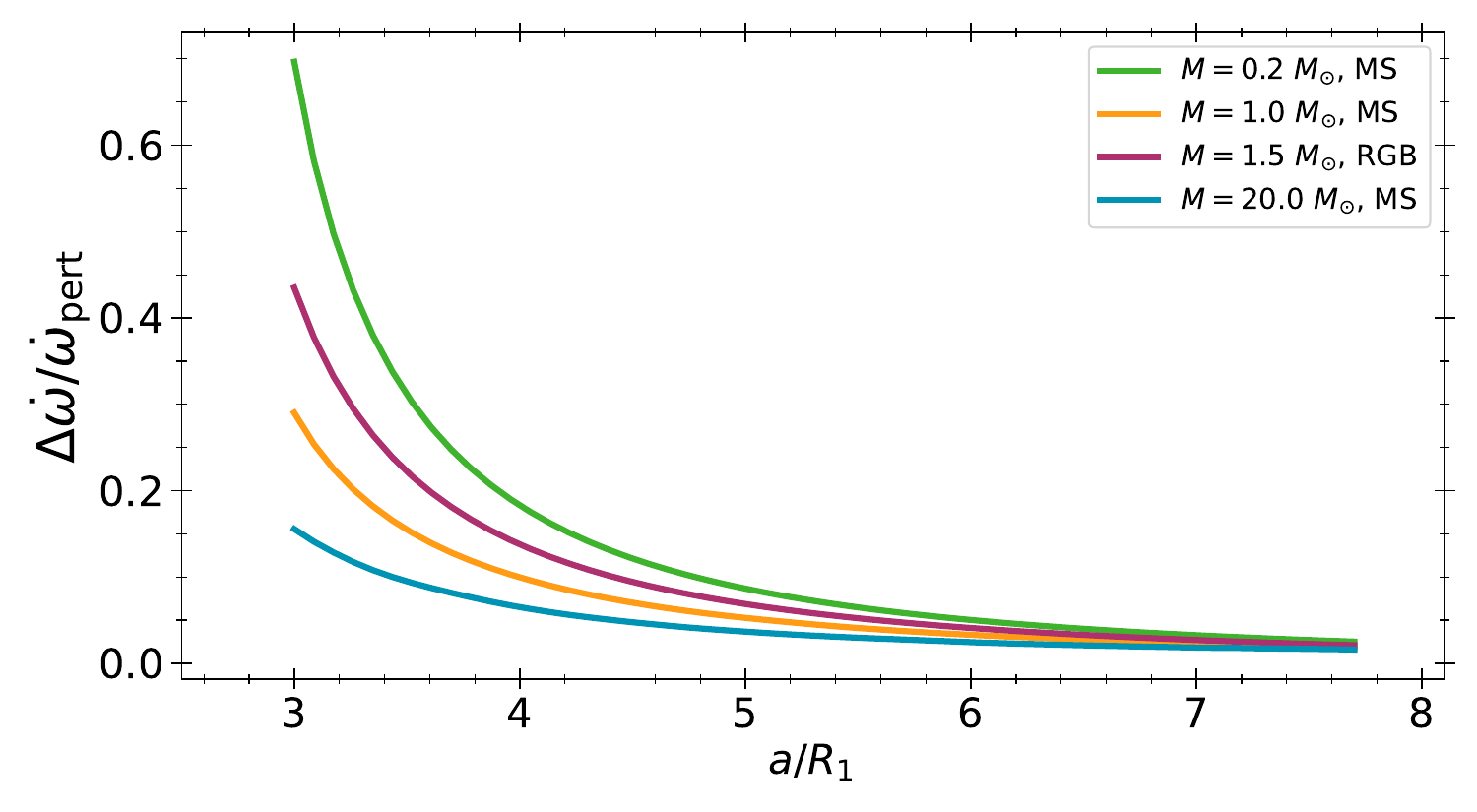}
\includegraphics[clip=true,trim=10 10 10 10,width=0.49\linewidth]{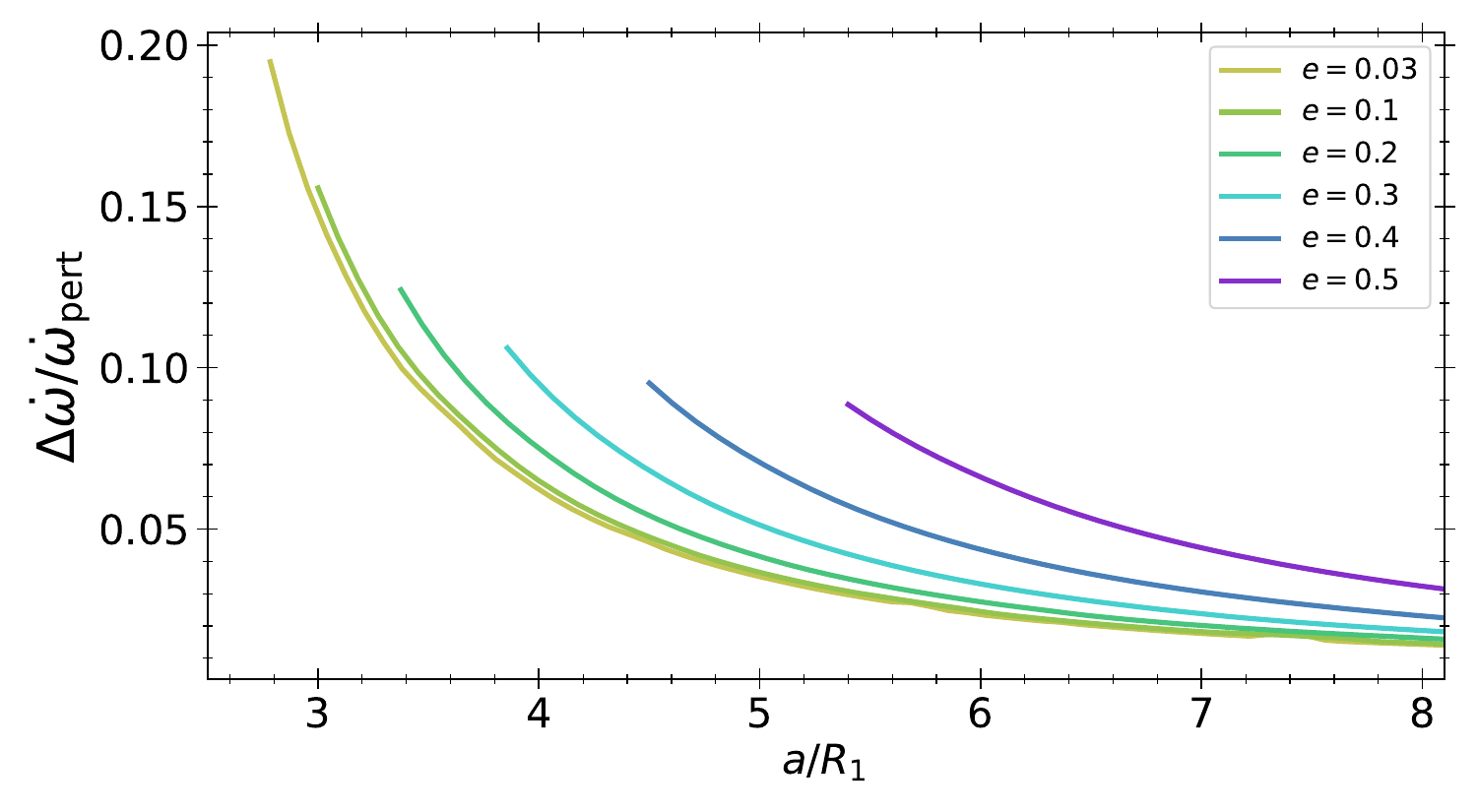}
\caption{Apsidal motion relative difference as a function of the orbital separation normalised by the stellar radii for the four same twin binaries as in Fig.\,\ref{fig:eta2} and an orbital eccentricity of 0.1 (\textit{left panel}), and for $20\,M_\odot$ main-sequence stars but different values of the eccentricity (\textit{right panel}). Figure adapted from \citet{fellay24}. }
\label{fig:mobidict}
\end{figure}

\citet{fellay23} demonstrated that the perturbative models significantly underestimate the deformations of binaries (see Fig.\,\ref{fig:eta2}). The impact is more pronounced for stars having a significant envelope mass compared to their total mass (low-mass main-sequence and red giant branch stars) and stars belonging to close binaries, when the orbital separation is comparable to the radii of the stars. Consequently, the impact on the apsidal motion rate follows the impact on $\eta_2$. Up to 70\% and 40\% errors are made on the apsidal motion rate determination for low-mass and red giant binaries at low-separation ($a/R_1 \sim 3$) and eccentricity (see Fig.\,\ref{fig:mobidict}, left panel). An even higher discrepancy is observed at a given orbital separation but higher eccentricity: The stars get closer during periastron passage, so are the stellar deformations more pronounced and the impact on the apsidal motion rate more important (see Fig.\,\ref{fig:mobidict}, right panel).

\citet{fellay24} applied the \texttt{MoBiDICT} code to four well-known twin binaries (PV\,Cas, IM\,Per, Y\,Cyg, and HD\,152248) to compare the parameters obtained to those from 1D stellar modelling. Overshooting was included in the models as the only mixing process. Except in the most massive binary HD\,152248, $\alpha_\text{ov}$ is systematically higher in the \texttt{MoBiDICT} model than in the 1D model. This is a direct consequence of the more important deformations in the non-perturbative approach and the ensuing smaller $k_2$. It requires even more mixing in the models to reproduce the $k_2$ \citep[see][their Table 2]{fellay24}. On a side note, the non-perturbative approach, though a small correction to the perturbative approach, is on the same order of magnitude as the general relativistic correction \citep[see][their Table 3]{fellay24}, supporting that it cannot be neglected anymore in the study of the apsidal motion in close eccentric binaries. 

\section{Apsidal motion as a means to estimate the stellar masses}
The previous sections have shown us that the double-line spectroscopic, eclipsing binaries are the most promising binaries as they allow us to derive all stellar and orbital parameters in a consistent way. In case the binary is only eclipsing, its lightcurves can turn out to be sufficient to derive the stellar and orbital parameters of the stars, including the apsidal motion rate. But what if the binary is a double-line spectroscopic but non-eclipsing binary? In that case, we indeed have no estimate of the masses and radii of the stars. Surprisingly, the apsidal motion equations combined with the binary' spectroscopic observations allow us to derive the masses of the stars, in a model-dependent way. 
 
Let's come back to the apsidal motion equations \eqref{eqn:omegadot}, \eqref{eqn:omegadotN}, and \eqref{eqn:omegadotGR}: They can be solved for the primary stellar mass $m_1$ if values for the stellar radii and $k_2$ are assumed from stellar models. This original method was first introduced by \citet[to which R. Barb\'a has contributed]{benvenuto02}. The principle rests on the sensible assumption that both stars have the same age and uses the known $q = m_2/m_1$ to get $m_2$ as a function of $m_1$. The radii and $k_2$ are derived from evolutionary calculations as a function of $m_1$ and the age. In that sense, the method is model- and age-dependent, but yet can be very reliable. The principle is the following: 
\begin{enumerate}
\item Compute grids of evolutionary models; 
\item Construct isochrones starting at Zero Age Main Sequence (ZAMS); 
\item For each isochrone, find the solution of the apsidal motion equations. The only independent quantity is $m_1$, hence we get a value of $m_1$ corresponding to the age of the isochrone. 
\end{enumerate} 
Two constraints have to be taken into account in this method: 1) The mass function imposes a minimum value for $m_1$ and 2) the system is detached, hence $R_1 + R_2 < a$. When using this method though, we must be very careful: Very accurate stellar models are required as $\dot\omega_\text{N}$ is highly sensitive to the radius and evolves as $R^5$; the method is thus also code dependent. 

\begin{figure}[t!]
\centering
\includegraphics[clip=true,trim=2 2 0 10,width=0.45\linewidth]{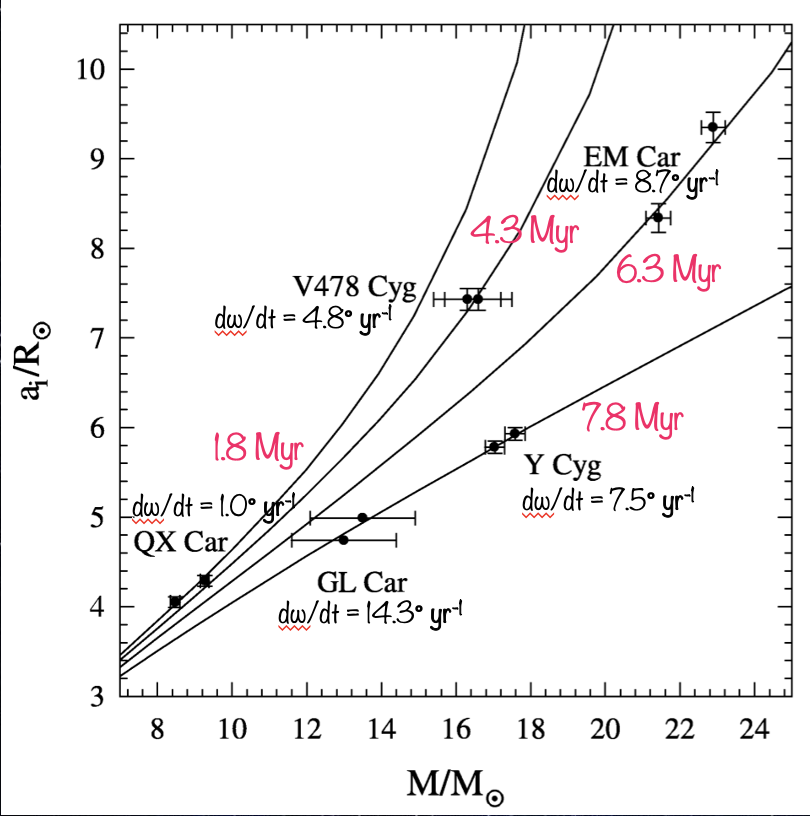}
\includegraphics[clip=true,trim=10 0 20 10,width=0.54\linewidth]{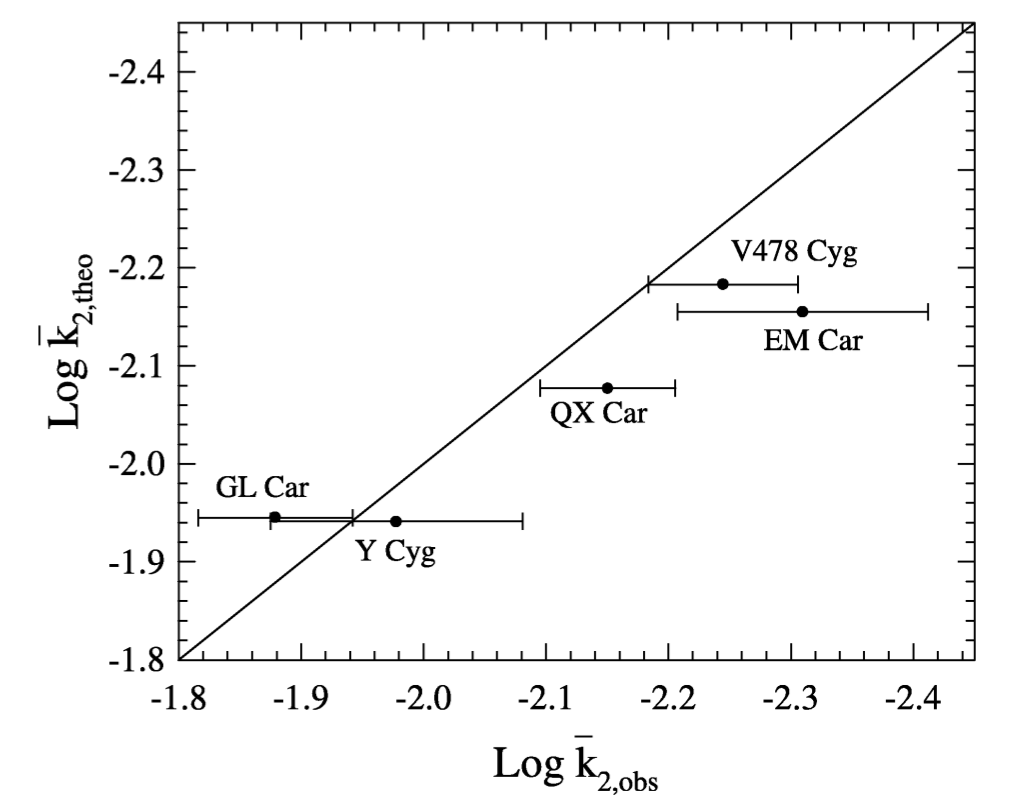}
\caption{\textit{Left panel:} Radius versus mass for five binary stars. Solid lines represent theoretical isochrones. \textit{Right panel:} Theoretical versus observational $\bar{k}_2$ for the same systems.  Figures adapted with permission from \citet{benvenuto02}.}
\label{fig:benvenuto1}
\end{figure}

To demonstrate the robustness of their method, \citet{benvenuto02} validated it against well-known eclipsing binaries for which the masses and radii are accurately determined. For each considered system, there is one isochrone that fits both components (see Fig.\,\ref{fig:benvenuto1}, left panel). The $k_2$ discrepancy appears clearly for QX\,Car, EM\,Car, and V478\,Cyg (see Fig.\,\ref{fig:benvenuto1}, right panel). 

\begin{figure}[h!]
\centering
\includegraphics[clip=true,trim=10 5 30 20,width=0.49\linewidth]{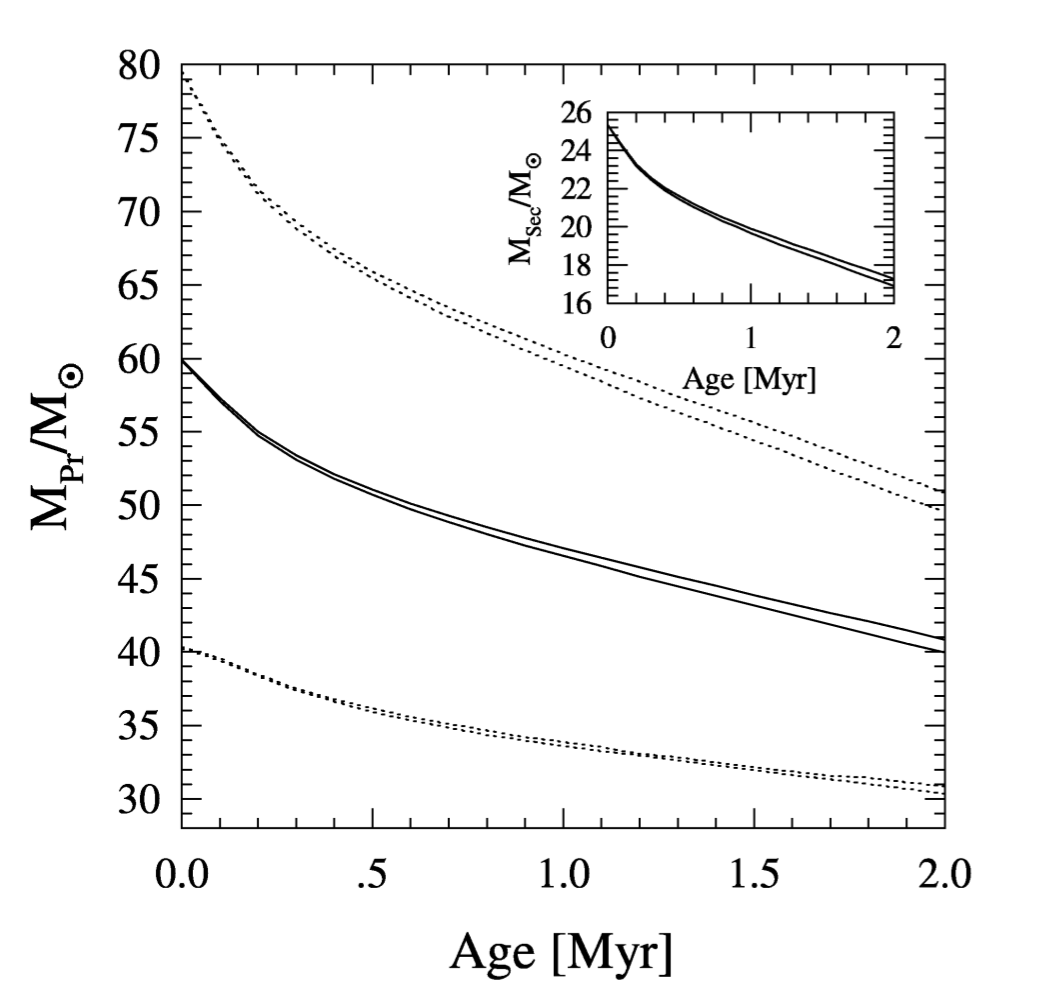}
\includegraphics[clip=true,trim=20 10 40 20,width=0.49\linewidth]{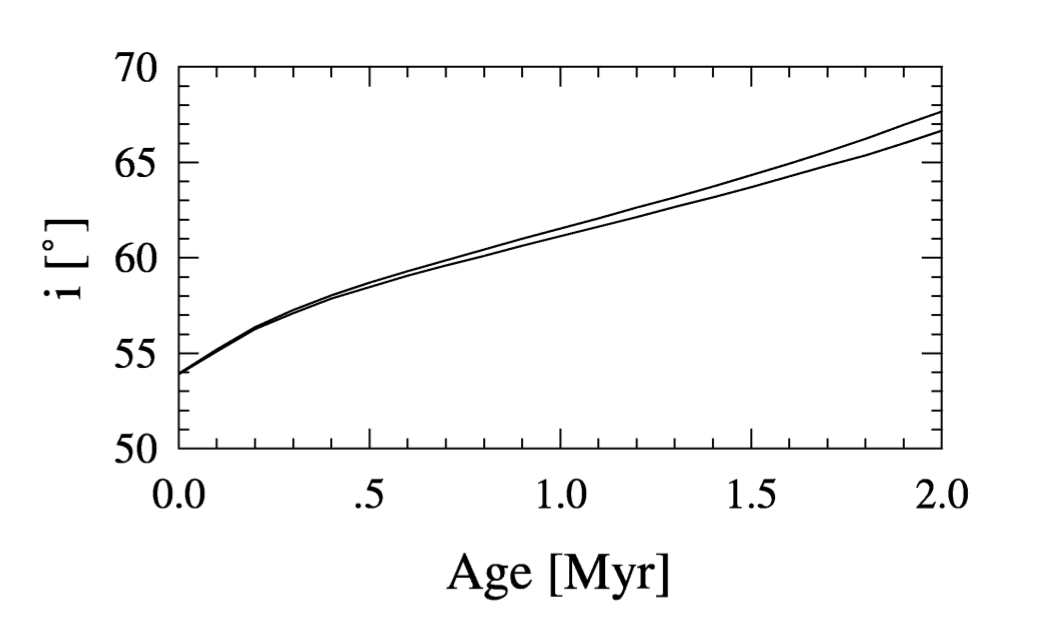}
\caption{\textit{Left panel:} Primary star's mass versus age of HD\,93205. The insert shows the secondary star's mass. The upper and lower solid lines correspond to the preferred value assuming $\alpha_\text{ov} = 0.40$ and 0.25, respectively. The dotted lines represent the $1\sigma$ uncertainties. \textit{Right panel:} Inclination of HD\,93205 versus age for the two values of $\alpha_\text{ov}$.  Figures taken with permission from \citet{benvenuto02}.}
\label{fig:benvenuto2}
\end{figure}

\citet{benvenuto02} applied their method to the massive non-eclipsing binary HD\,93205 made of an O3\,V primary and an O8\,V secondary. Its location in the open cluster Trumpler 16 in the Carina Nebula makes it difficult to establish its age with certainty, as there is ongoing star formation in the nebula. Yet, most massive stars of Trumpler 16 have an age ranging between 1 and 2\,Myr. Assuming an age of 0\,Myr or 2\,Myr, the authors got $m_1 = 60\pm19$\,M$_\odot$ and $m_2 = 25\pm8$\,M$_\odot$ or $m_1 = 40\pm9$\,M$_\odot$ and $m_2 = 17\pm4$\,M$_\odot$ (see Fig.\,\ref{fig:benvenuto2}, left panel). The main source of uncertainty in the masses determination comes from the apsidal motion. Yet, the masses of the secondary star are in agreement with expectations for O8\,V type stars. Based on their derived masses and the minimum masses determined from the RV adjustment, the authors determined that the inclination ranges between $54^\circ$ and $68^\circ$ (see Fig.\,\ref{fig:benvenuto2}, right panel), in agreement with the most probable value of $60^\circ$ from \citet{antokhina00}.

\begin{figure}[h!]
\centering
\includegraphics[clip=true,trim=10 0 10 10,width=0.36\linewidth]{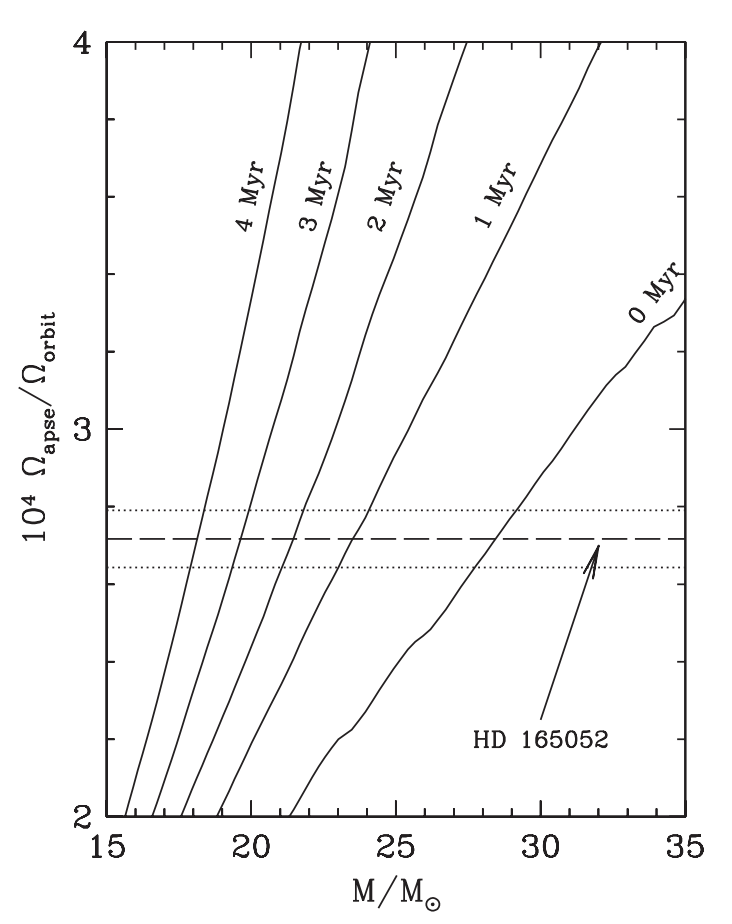}
\includegraphics[clip=true,trim=10 0 10 10,width=0.38\linewidth]{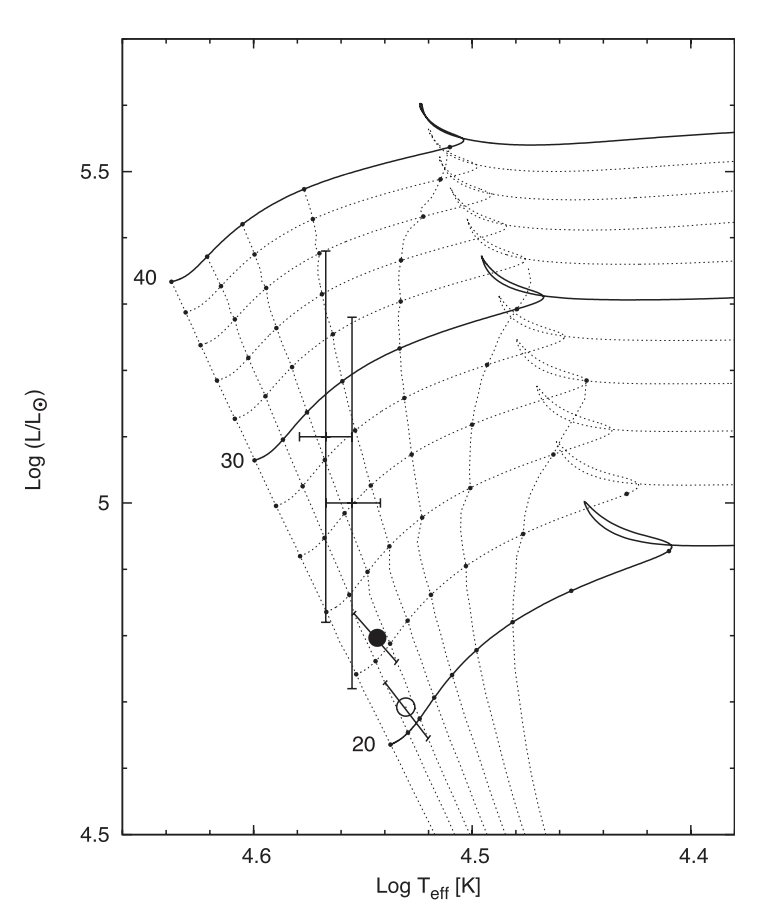}
\caption{\textit{Left panel:} Apsidal motion versus mass of the primary star of HD\,165052 for different values of the age of the system (solid lines). The observational apsidal motion rate and its error bars are depicted by the horizontal dashed and dotted lines, respectively. \textit{Right panel:} Evolutionary tracks in the HR diagram for stars of 20 to 40\,M$_\odot$. On each track, filled dots indicate time intervals of 1\,Myr. Locations of the stars according to the literature (crosses): Effective temperature dispersion in the observational calibration of \citet[Table 4]{martins05}, luminosity dispersion from distance modulus from \citet[lower limit]{mayne08} and $M_V$ from \citet[upper limit]{buscombe69}. Locations of the stars derived from the apsidal motion rate (primary and secondary stars depicted with filled and open circles, respectively). The error bars depict the age uncertainty propagated to masses. Figures taken with permission from \citet{ferrero13}.}
\label{fig:ferrero1}
\end{figure}

\citet{arias02} applied the same method to the massive binary HD\,165052 made of an O6.5\,V((f)) primary and an O7\,V((f)) secondary for which they detected the apsidal motion for the first time. Unfortunately, the authors could not derive consistent masses because of the hypothesis of synchronisation they had made. \citet{ferrero13} derived an apsidal motion rate of $12.1\pm0.3^\circ$\,yr$^{-1}$ for the system. They assumed an age of $1.5\pm0.5$\,Myr for the binary and derived masses $m_1 = 22.5\pm1.0$\,M$_\odot$ and $m_2=20.5\pm0.9$\,M$_\odot$ (see Fig.\,\ref{fig:ferrero1}, left panel), compatible with photometric masses of similar O-type stars and theoretical masses from \citet{martins05} calibration. Based on their derived masses and the minimum masses determined from the RV adjustment, the authors derived an inclination of $23^\circ$ for the system. However, the authors found a discrepancy between the location in the HR diagram as coming from the literature and the one derived from the apsidal motion (see Fig.\,\ref{fig:ferrero1}, right panel). The authors attributed this discrepancy to the difficulty of determining the cluster distance with accuracy.

\begin{figure}[h!]
\centering
\includegraphics[clip=true,trim=130 110 360 180,width=0.49\linewidth]{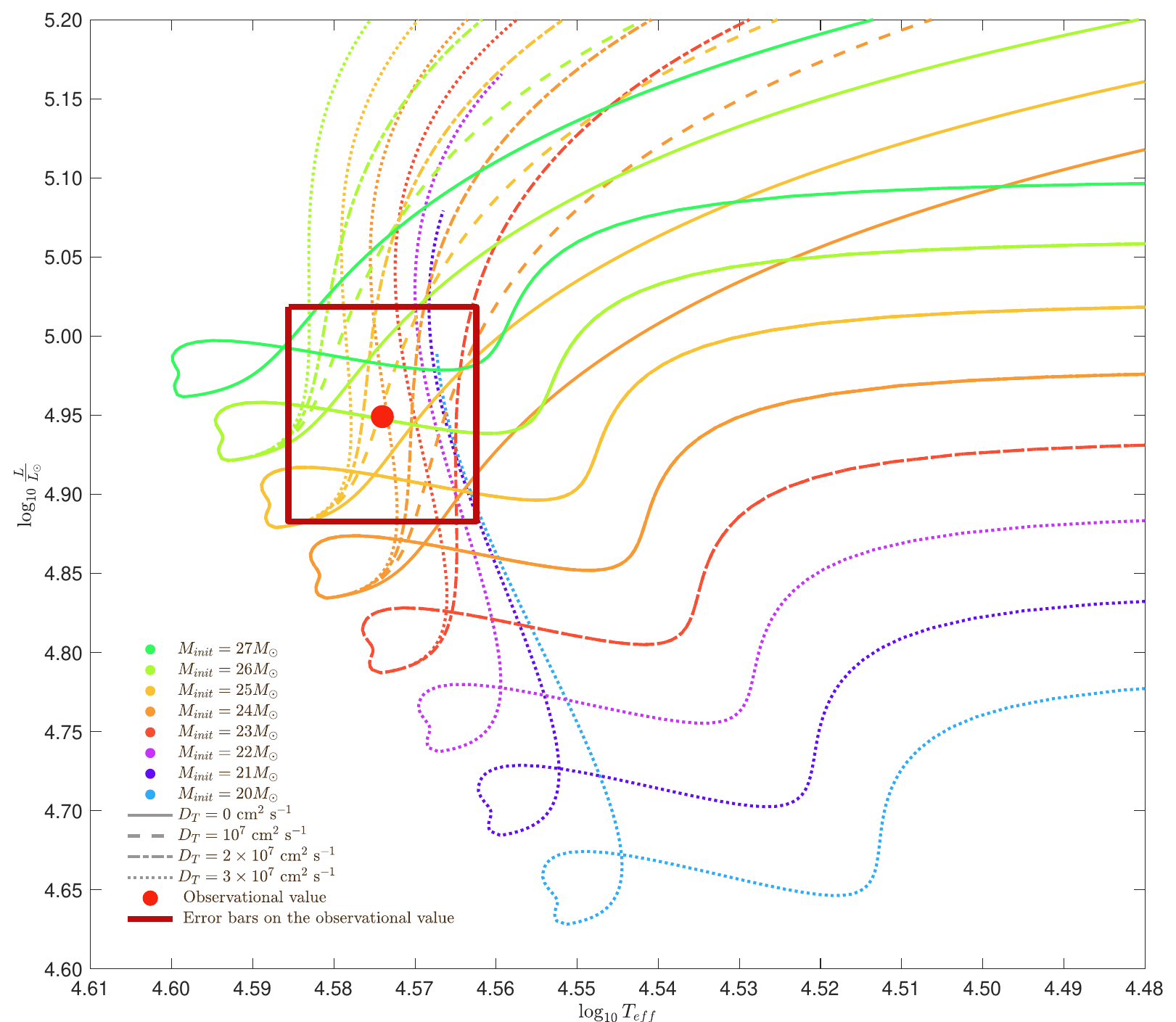}
\includegraphics[clip=true,trim=130 110 360 180,width=0.49\linewidth]{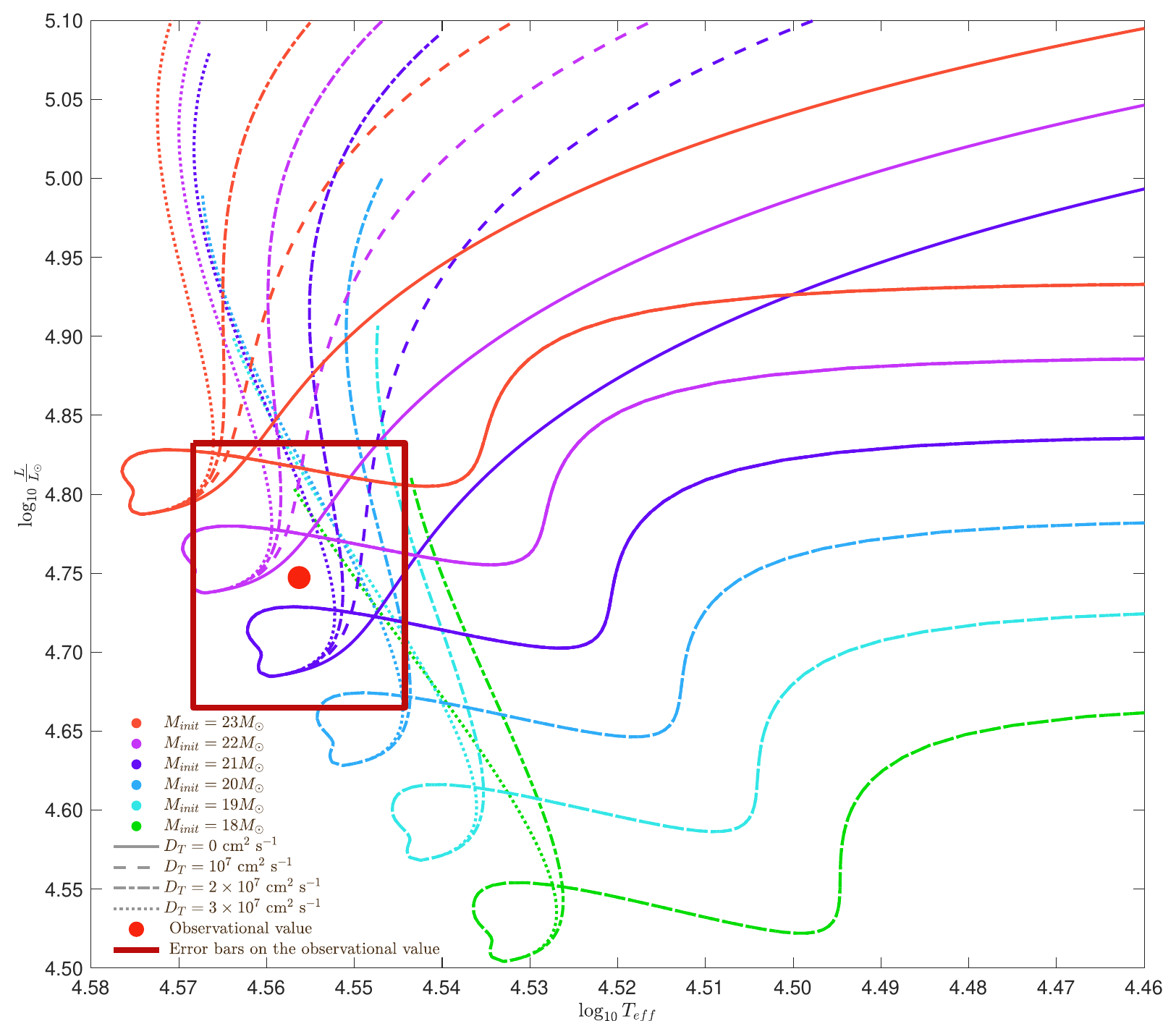}
\caption{HR diagrams for the primary (\textit{left panel}) and secondary (\textit{right panel}) star of HD\,165052: evolutionary tracks of \texttt{Cl\'es} models with different values of the initial mass and $D_T$. The observational value is represented by the red point, and its error bars are represented by the dark red rectangle. Figures taken from \citet{rosu23}.}
\label{fig:HD165052}
\end{figure}

We have revisited HD\,165052 and derived an apsidal motion rate of $11.3\pm0.6^\circ$\,yr$^{-1}$ \citep{rosu23}. We assumed an age of $2.0\pm0.5$\,Myr for the system and built several  \texttt{Cl\'es} evolutionary tracks with different initial masses and values of $D_T$ (see Fig.\,\ref{fig:HD165052}). From the analyses performed in \citet{rosu20a, rosu22a, rosu22b}, we estimated that $D_T = 2\times10^7$\,cm$^2$\,s$^{-1}$ should be the most appropriate value for this range of masses. Best-fit initial masses for the primary and secondary stars are of 25\,M$_\odot$ and 21\,M$_\odot$, respectively (see yellow dotted-dashed track in the left panel and dark purple dotted-dashed line in the right panel of Fig.\,\ref{fig:HD165052}). We derived $m_1 = 24.8\pm1.0$\,M$_\odot$ and $m_2 = 20.9\pm1.0$\,M$_\odot$. The error bars account for the error bars on the age, an error of 1\,M$_\odot$ on the initial mass, and an error of $10^7$\,cm$^2$\,s$^{-1}$ on $D_T$. Yet, the two stars are very young, so the value of $D_T$ does not have significant impact on the results as the mixing has not have time to significantly act yet. Based on our derived masses and the minimum masses determined from our RV adjustment, we constrained the inclination very accurately: $22.1^\circ < i < 23.3^\circ$ \citep{rosu23}. This range of values for the inclination is compatible with the value estimated by \citet{ferrero13}.

\begin{table*}[h!]
\caption{Theoretical values of the apsidal motion rate (in $^\circ$\,yr$^{-1}$) in HD\,165052. The values are obtained with models for the primary and secondary stars of 2.0\,Myr and the error bars are computed using models of 1.5 and 2.5\,Myr. In ``\textit{x}M\textit{yy}DT\textit{z}'', $x =$ P, S for the primary and secondary star, respectively, $yy$ is the initial mass of the model in M$_\odot$, and $z$ is the $D_T$ value in $10^7$\,cm$^2$\,s$^{-1}$.}
\label{table:HD165052}
\centering
\begin{tabular}{l | l l l l l l l}
\hline\hline
& \tiny{SM20DT3} & \tiny{SM21DT1} & \tiny{SM21DT2} & \tiny{SM21DT3} & \tiny{SM22DT1} & \tiny{SM22DT2} & \tiny{SM22DT3} \\ 
\hline
\vspace*{-5mm}\\
\tiny{PM24DT1} & \tiny{$10.54^{+0.34}_{-0.36}$} & \tiny{$11.13^{+0.43}_{-0.45}$} & \tiny{$11.00^{+0.38}_{-0.40}$} & \tiny{$10.92^{+0.35}_{-0.38}$} & \tiny{$11.52^{+0.47}_{-0.46}$} & \tiny{$11.38^{+0.41}_{-0.41}$} & \tiny{$11.29^{+0.37}_{-0.39}$} \\
\vspace*{-5mm}\\
\tiny{PM24DT2} & \tiny{$10.41^{+0.28}_{-0.31}$} & \tiny{$10.99^{+0.37}_{-0.40}$} & \tiny{$10.87^{+0.32}_{-0.36}$} & \tiny{$10.78^{+0.29}_{-0.33}$} & \tiny{$11.39^{+0.41}_{-0.41}$} & \tiny{$11.25^{+0.34}_{-0.37}$} & \tiny{$11.15^{+0.31}_{-0.34}$} \\
\vspace*{-5mm}\\
\tiny{PM24DT3} & \tiny{$10.32^{+0.25}_{-0.28}$} & \tiny{$10.90^{+0.34}_{-0.37}$} & \tiny{$10.78^{+0.28}_{-0.33}$} & \tiny{$10.69^{+0.26}_{-0.30}$} & \tiny{$11.30^{+0.37}_{-0.38}$} & \tiny{$11.15^{+0.31}_{-0.34}$} & \tiny{$11.06^{+0.28}_{-0.31}$} \\
\vspace*{-5mm}\\
\tiny{PM25DT1} & \tiny{$10.85^{+0.38}_{-0.38}$} & \tiny{$11.44^{+0.46}_{-0.47}$} & \tiny{$11.31^{+0.41}_{-0.42}$} & \tiny{$11.22^{+0.38}_{-0.40}$} & \tiny{$11.83^{+0.50}_{-0.48}$} & \tiny{$11.69^{+0.44}_{-0.43}$} & \tiny{$11.59^{+0.41}_{-0.41}$} \\
\vspace*{-5mm}\\
\tiny{PM25DT2} & \tiny{$10.71^{+0.31}_{-0.32}$} & \tiny{$11.29^{+0.39}_{-0.41}$} & \tiny{$11.16^{+0.34}_{-0.37}$} & \tiny{$11.08^{+0.31}_{-0.35}$} & \tiny{$11.68^{+0.43}_{-0.43}$} & \tiny{$11.54^{+0.37}_{-0.38}$} & \tiny{$11.45^{+0.34}_{-0.35}$} \\
\vspace*{-5mm}\\
\tiny{PM25DT3} & \tiny{$10.61^{+0.26}_{-0.29}$} & \tiny{$11.20^{+0.35}_{-0.38}$} & \tiny{$11.07^{+0.30}_{-0.34}$} & \tiny{$10.98^{+0.27}_{-0.32}$} & \tiny{$11.59^{+0.38}_{-0.40}$} & \tiny{$11.45^{+0.32}_{-0.35}$} & \tiny{$11.35^{+0.29}_{-0.32}$} \\
\vspace*{-5mm}\\
\tiny{PM26DT1} & \tiny{$11.18^{+0.40}_{-0.40}$} & \tiny{$11.76^{+0.49}_{-0.49}$} & \tiny{$11.63^{+0.44}_{-0.45}$} & \tiny{$11.54^{+0.41}_{-0.42}$} & \tiny{$12.14^{+0.52}_{-0.50}$} & \tiny{$12.00^{+0.46}_{-0.46}$} & \tiny{$11.91^{+0.43}_{-0.43}$} \\
\vspace*{-5mm}\\
\tiny{PM26DT2} & \tiny{$11.02^{+0.33}_{-0.35}$} & \tiny{$11.60^{+0.41}_{-0.44}$} & \tiny{$11.47^{+0.36}_{-0.39}$} & \tiny{$11.39^{+0.33}_{-0.37}$} & \tiny{$11.99^{+0.45}_{-0.45}$} & \tiny{$11.84^{+0.39}_{-0.40}$} & \tiny{$11.75^{+0.35}_{-0.37}$} \\
\vspace*{-5mm}\\
\tiny{PM26DT3} & \tiny{$10.91^{+0.28}_{-0.31}$} & \tiny{$11.49^{+0.37}_{-0.40}$} & \tiny{$11.36^{+0.32}_{-0.35}$} & \tiny{$11.28^{+0.29}_{-0.33}$} & \tiny{$11.88^{+0.40}_{-0.41}$} & \tiny{$11.74^{+0.34}_{-0.36}$} & \tiny{$11.64^{+0.31}_{-0.34}$}\\
\hline
\end{tabular}
\end{table*}

We performed all combinations of primary and secondary models accounting for the error bars on the initial mass and $D_T$ to compute theoretical apsidal motion rates (see Table\,\ref{table:HD165052}). Within the error bars, all combinations give an apsidal motion rate compatible with the observational value of $11.3\pm0.6^\circ$\,yr$^{-1}$, the closest models being those with initial masses of 25\,M$_\odot$ and 21\,M$_\odot$ for the primary and secondary stars, respectively, and $D_T = 2\times10^7$\,cm$^2$\,s$^{-1}$ for both stars, supporting our results.

\section{The apsidal motion technique to sound stellar interiors still in its infancy}
The observational determination of the apsidal motion rate in double-line spectroscopic, eclipsing, eccentric massive binaries is a beautiful method that uses Newton equations to probe the stellar interiors. The combined analysis of the radial velocities and lightcurves of one binary allows us to derive consistent and accurate stellar and orbital parameters for the system. The confrontation between stellar evolution models and the observational parameters allows us to constrain the inner density profiles of the stars and, hence, obtain constraints on the internal mixing processes occurring inside the stars. 

I have shown several studies demonstrating that the standard 1D stellar evolution models predict stars having a smaller internal stellar structure constant, that is to say, stars having a smaller density contrast, than expected from observations. I demonstrated with HD\,152248 and HD\,152219 as guinea pigs, that the addition of mixing inside the models helps to solve, at least partially, this discrepancy. Whether this additional mixing might be fully explained by rotationally-induced mixing is under investigation with \texttt{GENEC}. Studies with the non-perturbative code \texttt{MoBiDICT} showed that the perturbative model assumption is not justified in highly distorted stars, in which cases the apsidal motion is underestimated, exacerbating even more the need for enhanced mixing inside the models.

Using the apsidal motion to determine the stellar masses of a double-line spectroscopic but non-eclipsing binary is a method that goes off the beaten track that R. Barb\'a contributed to develop. Though this method is model-dependent, or more precisely, age- and mass-dependent, it turns out to be very accurate to derive both the stellar masses and orbital inclination, as I showed for HD\,93205 and HD\,165052.

Still too few massive binaries have now benefited from the analysis of their apsidal motion. It is of utmost importance to drastically increase our sample of binaries analysed this way and unveil the physical origin of the enhanced mixing we discovered.

% --------
% --------
% --------
% --------

\section*{Acknowledgements} 
I warmly thank Jesús Maíz Apellániz for having invited me to this friendly and enriching conference in honour of Rodolfo Barbá.

\bibliographystyle{ceab}
\bibliography{apsidal_motion}

\end{document}